\documentclass[trackchanges, twocolumn]{aastex7}
\usepackage{amsmath}
\usepackage{multirow}
\usepackage{graphicx}
\usepackage{longtable}
\usepackage{adjustbox}
\usepackage{threeparttable}
\usepackage{gensymb} 
\usepackage{comment}
\usepackage{overpic}

\usepackage{tikz}
\usetikzlibrary{calc,tikzmark} 
\newlength{\imageheight}
\begin{document}

\title{Characterizing the velocity anisotropy of the Milky Way's stellar halo}

\author[0000-0002-8262-2246]{F.~O.~Barbosa}
\affiliation{Universidade de S\~ao Paulo, Instituto de Astronomia, Geof\'isica e Ci\^encias Atmosf\'ericas, Departamento de Astronomia, SP 05508-090, S\~ao Paulo, Brasil}
\email[show]{fabriciaob@usp.br}

\author[0000-0002-5974-3998]{A. P\'erez-Villegas} 
\affiliation{Instituto de Astronom\'ia, Universidad Nacional Aut\'onoma de M\'exico, Apartado Postal 106, C. P. 22800, Ensenada, B. C., M\'exico}
\email{mperez@astro.unam.mx}

\author[0000-0001-7479-5756]{S. Rossi}
\affiliation{Universidade de S\~ao Paulo, Instituto de Astronomia, Geof\'isica e Ci\^encias Atmosf\'ericas, Departamento de Astronomia, SP 05508-090, S\~ao Paulo, Brasil}
\email{silvia.rossi@iag.usp.br}

\author[0000-0002-7529-1442]{R. M. Santucci}
\affiliation{Universidade Federal de Goi\'as, Campus Samambaia, Instituto de F\'isica, Goi\^ania, GO 74001-970, Brasil}
\affiliation{Universidade Federal de Goi\'as, Instituto de Estudos Socioambientais, Planet\'ario, Goi\^ania, GO 74055-140, Brasil}
\email{rafaelsantucci@ufg.br}

\author[0000-0001-7336-2836]{L. Aguilar} 
\affiliation{Instituto de Astronom\'ia, Universidad Nacional Aut\'onoma de M\'exico, Apartado Postal 106, C. P. 22800, Ensenada, B. C., M\'exico}
\email{aguilar@astro.unam.mx}

\author[0000-0002-0537-4146]{H. D. Perottoni}
\affiliation{Observat\'{o}rio Nacional, MCTI, Rua Gal. Jos\'{e} Cristino 77, Rio de Janeiro, 20921-400, RJ, Brasil}
\email{hperottoni@on.br}

\author[0000-0002-9269-8287]{G. Limberg}
\affiliation{Kavli Institute for Cosmological Physics, University of Chicago, 5640 S Ellis Avenue, Chicago, IL 60637, USA} 
\affiliation{Department of Astronomy \& Astrophysics, University of Chicago, 5640 S Ellis Avenue, Chicago, IL 60637, USA}
\email{limberg@uchicago.edu}

\author[0000-0003-3382-1051]{L. Borbolato}
\affiliation{Universidade de S\~ao Paulo, Instituto de Astronomia, Geof\'isica e Ci\^encias Atmosf\'ericas, Departamento de Astronomia, SP 05508-090, S\~ao Paulo, Brasil}
\email{laisborbolato@usp.br}

\author[0009-0007-5867-0583]{J. V. Nogueira-Santos}
\affiliation{Instituto Nacional de Pesquisas Espaciais, SP 12227-010, S\~ao Jos\'e dos Campos, Brasil}
\email{nogueiranogueirajoao@usp.br}

\begin{abstract}

Modeling the Milky Way stellar halo requires well-determined density and velocity anisotropy profiles. However, it has been challenging to gather a large sample of stars with six-dimensional data that extend beyond 40 kpc to map the outer halo. 
Our work investigates the velocity anisotropy in the Milky Way stellar halo with more than 10,000 blue horizontal-branch stars, combining Gaia astrometric data and spectroscopic data from SEGUE, DESI and LAMOST. This large sample allows us to obtain a detailed profile of up to $\sim 50$ kpc. 
Radial velocities are predominant in the inner halo ($< 30$ kpc), and the anisotropy presents a smooth decrease before rapidly dropping to negative values, becoming dominated by tangential dispersion velocities. 
Removing the main known accreted structures of the Milky Way, makes the anisotropy profile radially-dominated at all radii. Our profile clearly shows an increase in the anisotropy from the center of the Galaxy, in accordance to the simulations. 
We also investigate the correlation of anisotropy with metallicity and with color. The lack of correlation between metallicity and anisotropy in our clean sample reinforces that this relation is driven by merger events. 
The initial exploration with color indicates a relation between kinematics and age, showing that older stars are dynamically colder and present less radial orbits than younger stars in the inner halo.

\end{abstract}



\section{Introduction} \label{sec:intro}

The current cosmological model states that galaxies form by the merging of smaller systems. This idea became one of the main scenarios for the Galactic halo formation after the work of \cite{Searle78}. However, for a long time, the Milky Way (MW) constituents remained hidden by the limited access to regions beyond the solar neighborhood. 
The first direct observation that would consolidate the hierarchical formation came with \citet[][the Sagittarius (Sgr) stream]{Ibata1994}. They detected what seemed to be a dwarf galaxy, characterized by a low-velocity dispersion and a mean line-of-sight velocity distinctive of field stars in the direction of the bulge. Following this, another building block would be discovered as a coherent group in angular momentum space by \cite{Helmi99}, known as the Helmi streams (str). 
Dynamical methods (mainly the one discussed in \citealt{Helmi2000}) would later become one of the most widely used techniques to identify moving groups in the halo. 

When astrometric data were only available for a small set of stars, spectroscopic surveys had a fundamental role in Galactic studies. Thanks to Sloan Digital Sky Survey \citep[SDSS;][]{sdss} data, differences between closer and more distant halo stars were detected, showing an inner halo more metal-rich and flatter than the outer halo.  To explain this distinction, \cite{Carollo2007} presented the dual halo hypothesis, proposing that each component had its own formation process. Besides the spatial and chemical differences, a break in the density profile also indicated that the MW halo was not a single homogeneous component (e.g., \citealt{Saha1985}, \citealt{Watkins2009}, \citealt{Sesar2011}, \citealt{Deason2011}, \citealt{Amarante2024}, \citealt{Medina2024}). 

Our knowledge about the formation history of the MW was highly improved by the arrival of unprecedented quality data gathered by the Gaia satellite \citep{Gaia_mission}. The first proper motion measurements combining Gaia and SDSS data confirmed that a metal-rich ([Fe/H] $\gtrsim -1$) structure lives in the inner halo of MW (the Gaia-Sausage/Enceladus -- GSE; \citealt{Belokurov2018}, \citealt{Haywood2018}, \citealt{Helmi18}). The available six-dimensional data also permitted the widespread use of orbital properties and clustering algorithms to identify many other structures that constitute our halo (\citealt{myeongStreamsAndClumps,myeongShards}, \citealt{yuan2020}, \citealt{Gudin2021}, \citealt{Limberg2021}, \citealt{RuizLara2022}, \citealt{Dodd2023}), with some works suggesting that the entire halo might consist of accreted systems (e.g., \citealt{Bell2008}, \citealt{naidu2020}, but see also \citealt{Davies2025} suggesting a lower fraction). 
Distance estimates independent of the stellar evolutionary stage have allowed precise kinematic studies beyond the solar neighborhood, improving the level of details of the inner halo mapping (\citealt{Deason2017}, see a recent review by \citealt{Hunt2025}). However, parallax uncertainties increase significantly for Gaia magnitude $\rm G > 17$, reducing the reliability of these measurements for faint stars \citep{Lindegren2021}.

The limited observations in the outer halo hindered the MW mass determination. 
Prior to Gaia data, kinematic anisotropy estimates were based on assumed distribution functions to model the stellar populations, or relied on a few nearby halo stars with measured proper motions (e.g., \citealt{Deason2012}, \citealt{Kafle2012}, \citealt{Kafle2014}, \citealt{Williams2015}, \citealt{Cunningham2016}). 
The values varied significantly among works, going from radially- to tangentially-dominated orbits. \cite{Wang2015} showed that adopting a distribution function for an incomplete velocity space biases the velocity anisotropy. Including proper motions is crucial to constrain the velocity anisotropy and reduce correlations with other halo properties.

Modeling a sample of blue horizontal-branch (BHB) stars, \cite{Sommer-Larsen1994} and \cite{Sommer-Larsen1997} noted a transition in the velocity anisotropy, with a decrease in radial velocity dispersions and an increase of the tangential component beyond 20 kpc from the Galactic center. The radially-biased anisotropy has been consistently found in studies mapping the halo within 10 kpc, with different values (\citealt{Pier1984}, \citealt{Chiba1998}, \citealt{Smith2009}, \citealt{Bond2010}). 
With the increase in astrometric data available, we have been able to access the full kinematics of halo stars, enabling a direct determination of the velocity anisotropy profile. 
The inner region ($< 20$ kpc) has been established as radially dominated (\citealt{Belokurov2018}, \citealt{Wegg2019}, \citealt{Bird2019, Bird2021}, \citealt{Liu2022}, \citealt{Wu2022}). 
However, estimates in the outer part have not yet converged.

Most of the halo needs to be mapped by giants, preferably standard candle stars, among which we find the BHB stars. BHB stars are hot ($\gtrsim 7500 \rm \ K$), typically old ($\gtrsim 10 \ \rm Gyr$; \citealt{Dotter2010}) and metal-poor ([Fe/H] $\lesssim -0.5$; \citealt{santucci_frac}), with a couple of advantages over other common tracers. Their high and almost constant absolute magnitude allows a more precise estimate of photometric distances compared to K giants (\citealt{xue08, xue11, Xue2014}, \citealt{Loebman2018}). Moreover, they do not require time series photometry necessary for variable stars, making observations less time-consuming. Compared to the red horizontal-branch stars, they are more easily distinguishable from red dwarfs, which are abundant at small distances. Unfortunately, few spectroscopic surveys have been dedicated to extracting information from such hot spectra. Still, the available line-of-sight velocity and metallicities, combined with astrometric data, can be used to analyze the structure of the stellar halo at large distances (beyond 60 kpc, \citealt{xue08}, \citealt{barbosa22}).

In this work, we combine spectroscopic samples of BHB stars to explore the stellar halo, from the inner regions to beyond 40 kpc, investigating the influence of the main building blocks currently known. 
We find a transition from radially to tangentially-biased anisotropy around 35 kpc as a result of the combination of accreted systems that constitute the MW halo. When these structures are removed, the halo is radially biased at all radii. 
This study presents a detailed anisotropy profile, confirming the dependence on metallicity caused by the different evolutionary histories of each system, and the increase in the anisotropy in the inner region as predicted by simulations.

\section{Data}
\label{sec:data}

In this work, we combined publicly available spectroscopic catalogues with observations of the stellar halo from $\sim \rm 5 \ kpc$ to $\gtrsim 40 \rm \ kpc$. 
Each sample was crossmatched with Gaia DR3 catalog \citep{GaiaDR3} using a radius of 5$''$ to obtain proper motions and photometry ($\rm G, \ G_{RP}$, and $\rm G_{BP}$).

\subsection{Catalogues}

The catalogs presented here have dedicated works that focus on selecting and analyzing BHB stars. 
Here, we summarize the procedures for each one.

\subsubsection{BHB stars in SDSS/SEGUE}
\label{segue}

SDSS was conducted using the Sloan Foundation and the Irénée du Pont 2.5-m telescopes located, respectively, at Apache Point Observatory and at Las Campanas Observatory. Up to the sixteenth data release \citep[DR16;][]{sdssdr16}, it had observed almost 1 billion objects in five broad-bands (\textit{u, g, r, i,} and \textit{z}). Low-resolution (R $\sim 1800$) spectroscopic data were provided by the Sloan Extension for Galactic Understanding and Exploration \citep[SEGUE;][]{yanny}, processed by the SEGUE Stellar Parameter Pipeline \citep[SSPP;][]{Lee08a,Lee08b}.

BHB stars with spectroscopic information from SEGUE analyzed here were made available by \cite{barbosa22}. They selected the A-type stars based on photometric data from SDSS DR16, applying color cuts  $-0.3 < (g-r)_0$\footnote{The subscript 0 indicates the color has been deredded by the extinction coefficients provided in the SDSS catalog.} $< 0.1$ and $0.8 < (u-g)_0 < 1.4$. These ranges exclude most other stellar objects, such as white dwarfs and cooler stars. Still, to minimize the contamination of non-A-type stars and poor-quality data, additional cuts of effective temperature ($7500\,{\rm K} < \rm T_{\rm eff} < 10000\,{\rm K}$) and signal-to-noise ($\rm S/N > 10$) were applied. 

Giant and dwarf A-type stars can be segregated by the surface gravity ($\rm \log g$) provided by the SSPP \citep{santucci_frac}. \cite{barbosa22} used a two-dimensional Gaussian Mixture Model with $\rm \log g$ estimates $\log \rm{g}_{\rm ANNRR}$ and $\log \rm{g}_{\rm SPEC}$. 
This method results in a sample of 5,699 stars with a probability of being a BHB star $> 0.5$.

\subsubsection{BHB stars in DESI}
\label{desi}

The Dark Energy Spectroscopic Instrument (DESI, \citealt{DESI_overview}) is a multi-object spectrograph at the 4-m Mayall Telescope mounted at Kitt Peak National Observatory. It has a resolution between 2000 and 5000, with the lowest values corresponding to the shortest wavelengths.  

The BHB sample in DESI was selected by \cite{Bystrom2025}. These stars were observed as primary and secondary targets (corresponding to bright and dark programs), reaching magnitudes between $16 < r < 21$ in the DESI Legacy Imaging Survey \citep{Dey19_decals} optical photometry. The initial target selection was cleaned by removing quasar candidates (\texttt{RR\_SPECTYPE = QSO}) and bad line-of-sight velocity estimates based on the flags \texttt{CHISQ\_WANR, RV\_WARN} and \texttt{RVERR\_WARN}. The presence of RR Lyrae (RRL) stars was inspected by crossmatching the sample with Gaia DR3 \citep{Clementini2023_gaia_RRL} and PanSTARRS1 \citep{Sesar2017_panstarrs_RRL} catalogues. Main sequence stars were removed by limiting the surface gravity in 
\begin{center}
$1.50 < \log \rm g - 1.4\cdot10^{-4} \ T_{eff} < 2.65$, 
\end{center}
with $7000 < \rm T_{eff} < 14500 \ K$. The colour cut $-0.3 < (g-r)_0 < 0.0$ was also imposed. With these criteria, they obtained a sample of 9,866 BHB stars, from which 6,327 are publicly available.

For the present work, duplicate targets were filtered by the \texttt{PRIMARY} flag, resulting in 5,436 stars. We removed sources with \texttt{RV\_WARN = 8}, which indicates the object reaches the lower edge of the metallicity range ($\rm [Fe/H] \approx -4$). Objects with median signal-to-noise between the three arms below 10 per pixel were also removed. Our final DESI sample contains 4,176 BHB stars.

\subsubsection{BHB stars in LAMOST}
\label{lamost}

The Large Sky Area Multi-Object Fiber Spectroscopic Telescope (LAMOST; \citealt{Cui2012_LAMOST}, \citealt{Luo2012_LAMOST}, \citealt{Zhao2012_LAMOST}) located in Xinglong Station of National Astronomical Observatory, has provided low-resolution (R $\sim$ 1800) spectra for more than 11 million objects in the last public data release (DR10).

\cite{Ju2024} compiled a sample of A-type stars in the LAMOST DR5, based on the equivalent widths (EW) of the $\rm H_{\gamma}$ hydrogen line and G-band at 430 nm, using the limits: $\rm EW_{H_\gamma} > 8$ and $\rm EW_{G} < 2$, relying on the distribution presented by \cite{Liu2015}. 
To restrict their sample to metal-poor stars, they selected candidates based on $\rm EW_{Ca II K}$ and Gaia $\rm (G_{BP}-G_{RP})_0$ color space, adopting the expected values for a star with $\rm \log g = 3.5$ and $\rm [Fe/H] = -1$. 
In addition, they excluded stars with signal-to-noise in the $g$-band lower than 10, and those with Galactic latitude below $20\degree$ and above $-20\degree$. 

To select BHB stars, they combined the $f_m$ vs. $D_{0.2}$ method \citep{pier} and the scale-width--shape method \citep{Clewley_method_bc}. In the former, $f_m$ is the minimum flux relative to the continuum level and $D_{0.2}$ measures the line width of Balmer lines at 20\% of the continuum, providing indirect information about $\rm T_{\rm eff}$ and $\log \rm{g}$. On the other hand, the scale-width–-shape method describes the shape of the lines relying on the parameters of a Sérsic profile \citep{sersic}. 
To achieve a larger number of BHB stars with good precision, they employed the two methods with three hydrogen lines: $\rm H_{\gamma}$, $\rm H_{\delta}$ and $\rm H_{\beta}$. Combining the three groups, they found 5,355 BHB candidates. 

A following work \citep{Ju2025} estimated atmospheric parameters for those BHB stars using the Stellar Label Machine code trained with A-type synthetic spectra from \cite{AllendePrieto2018}. The grid of spectra was refined, interpolating the 3,200 spectra to generate 5000 spectra equally spaced in the range of $7,000 < \rm T_{eff} < 12,000 \ K$ and $2.5 < \rm \log g < 5.0$. 
This procedure provided metallicities and $\log \rm{g}$ for all the stars in the initial sample. 
To obtain the line-of-sight velocities, we crossmatched the catalog with LAMOST DR10, reducing the sample to 3,902 stars. Five stars were removed due to the absence of proper motion and/or $\rm G_{BP}/G_{RP}$ photometry from Gaia DR3, leaving us with 3,897 BHB stars.

\subsection{Creating the complete sample}

Combining data from SDSS/SEGUE, LAMOST and DESI, we obtain 12,655 unique BHB stars. Data from SDSS/SEGUE and DESI were prioritized when removing duplicated sources, as the former was used to anchor the photometry and LAMOST line-of-sight velocities possess larger uncertainties.
SDSS $g-$ and $r-$bands for DESI and LAMOST sources were obtained by transforming Gaia photometry, following \cite{gaia_phot}. The magnitudes were corrected using reddening from \cite{Schlegel1998} and the relative extinction from \cite{WangChen2019}. 
The transformation was verified with common stars between the catalogues (339 with LAMOST and 801 with DESI), and the measured offset was applied to both samples. 
Metallicity and line-of-sight velocity estimates were also verified to take into account any systematics between the surveys. The analysis performed and the respective offsets applied are presented in the Appendix \ref{ap:A}. 
With the calculated bands, absolute magnitudes were estimated using the color–magnitude relation from \cite{barbosa22} to derive distances.

Galactocentric coordinates were calculated using \texttt{astropy} \citep{astropy2013}. The corresponding Cartesian coordinates were obtained adopting Sun's position in a Galactocentric system as (X, Y, Z)$_{\odot} = (-8.2, 0.0, 0.0) \ \rm kpc$ (\citealt{Bland-Hawthorn2016}). 
Velocities and orbital parameters were calculated adopting solar velocity (U, V, W)$_{\odot} = (11.10, 12.24, 7.25) \ \rm{km\,s}^{-1}$ \citep{Schonrich10} and local standard of rest system velocity $\rm V_{LSR} = 232.8 \ \rm{km\,s}^{-1}$ \citep{McMillan17}. 
We opted to present the final results in a left-handed frame, with stars in prograde motion having angular momentum in the Galactic plane $L_z > 0$.

\begin{figure*}[t]
\centering
  \includegraphics[width=2\columnwidth]{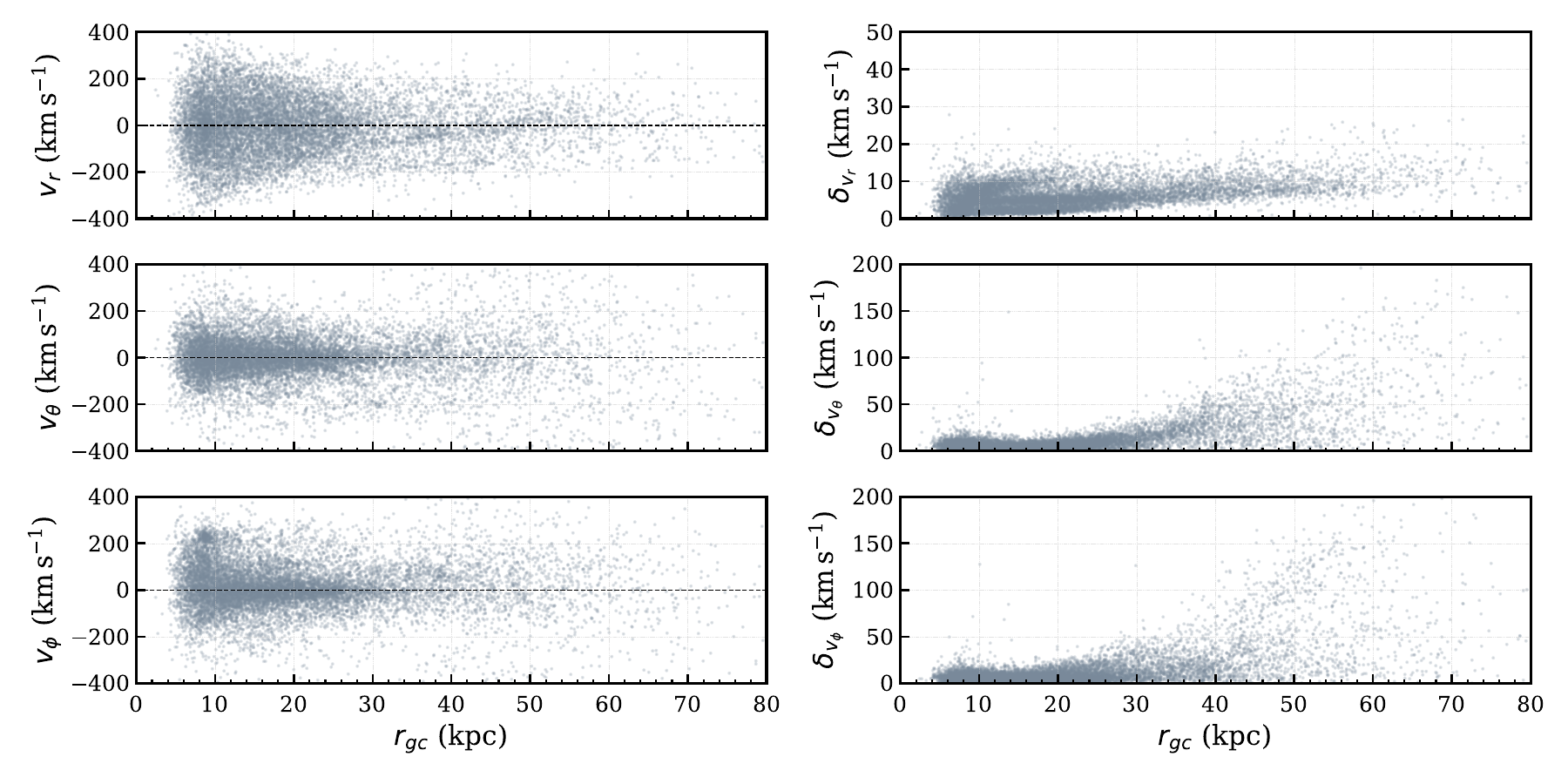}
  \caption{Left column: distribution of velocity components in spherical coordinates vs. Galactocentric distance for the assembled sample. Right column: velocity uncertainties vs. Galactocentric distance for the assembled sample.}
  \label{fig:veloc}
\end{figure*}

All kinematic parameters were adopted as the median values of 100 Monte Carlo (MC) realizations, taking into account the effect of the observational uncertainties (for $g$ and $r$ bands, proper motions, and line-of-sight velocities) and the error of the absolute magnitude relation \citep{barbosa22}. We adopted half the difference of the 16th and 84th percentiles from MC realizations as the uncertainties for estimated parameters.
In Figure \ref{fig:veloc}, we present radial ($v_r$), polar ($v_\theta$) and azimuthal ($v_\phi$) velocity components and corresponding uncertainties as a function of Galactocentric distance ($r_{gc}$) for the assembled sample. We can observe a group of stars with prograde motion $v_\phi \approx 200 \ \rm km\,s^{-1}$, which corresponds to the disk of the MW.

To avoid including main-sequence and blue straggler stars that may be included in the works that performed the selection, objects with $\rm \log g > 4.5$ and $\rm [Fe/H] > -1$ were removed. Moreover, stars outside $-0.3 < (g-r)_0 < 0.1$ calculated with magnitude transformation and \texttt{ruwe} $> 1.4$ were also excluded. 
Stars with velocity uncertainties larger than $100 \ \rm km \,s^{-1}$ for radial component, and $150 \ \rm km \,s^{-1}$ for azimuthal and polar components were excluded for the following analysis \citep{Bird2021}, as well as stars with uncertainties larger than 20\% in distance.

Since our main interest is analyzing the stellar halo kinematic, disk stars were removed after inspecting their orbital properties, such as perigalactic and apogalactic distances, eccentricity and maximum vertical excursion from the Galactic plane ($\rm Z_{max}$). The orbits were integrated using \texttt{AGAMA} package (\citealt{AGAMA}) during 13 Gyr in a \cite{McMillan17} Galactic potential, which includes a spherical dark matter halo, stellar and gaseous disks, and a bulge. MC simulation was also employed to estimate dynamical parameters.
We observed several stars with azimuthal velocity and eccentricity corresponding to thin/thick disc reaching vertical coordinate $\rm Z_{GC} \sim 8 \ kpc$. Therefore, we chose not to use $\rm Z_{GC}$ or $\rm Z_{max}$ and instead adopted as disc stars those with eccentricity $< 0.5$ and an angular momentum fraction $\rm L_z/L_{tot} > 0.9$.

The final halo sample is composed of 10,323 BHB stars. Hereafter, we refer to this sample as the complete sample. In Table \ref{tab:summary}, we summarise the main properties by survey.

\begin{table}[h]
\caption{Main properties for stars in the complete sample. Number of stars (N), Galactocentric distance ($\rm r_{gc}$), velocity components ($v_r$, $v_\theta$ and $v_\phi$), metallicity ([Fe/H]), and magnitude $g_0$ ranges are presented according to the source survey.} \label{tab:summary}
\hspace{-1.3cm}
\resizebox{9.7cm}{!}{
\centering
\begin{tabular}{lccc}
 Survey & SEGUE & LAMOST & DESI \\
\hline \hline 
N & 4900 & 2294 & 3129 \\
$\rm r_{gc} \ \rm (kpc)$  & [2.4, 86.2] & [3.8, 35.5] & [6.0, 114.6] \\
$v_r \ \rm (km \, s^{-1})$  & $[-377.5$, +391.5] & [$-403.2$, +408.4] & [$-332.0$, +360.2] \\
$v_\theta \ \rm (km \, s^{-1})$  & [$-509.9$, +441.7]  & [$-346.6$, +443.3] & [$-362.0$, +407.3] \\
$v_\phi \ \rm (km \, s^{-1})$   & [$-444.3$, +432.9] & [$-322.0$, +500.1] & [$-363.8$, +333.5] \\
$\rm[Fe/H]$ & $[-3.3, -1.0]$ & $[-3.3, -1.0]$ & $[-3.8, -1.0]$ \\
$g_0$ & $[13.4, 20.1]$ & $[9.4, 17.7]$ & $[15.7, 20.6]$ \\
\hline \hline 
\end{tabular}
}
\end{table}

\section{Structures selection}
\label{sec:strucs}

Given their high effective temperatures, BHB stars lack reliable $\alpha$ abundance estimates in large surveys such as DESI and LAMOST, hindering an identification of accreted structures based on their chemical composition. 
Using mainly spatial distribution and orbital parameters, \cite{horta23} could identify members of important accretions and confirm their extragalactic origin with APOGEE chemical information \citep{apogee}. 
Following this strategy, we adopted the kinematic and dynamic criteria presented in the literature to segregate the components of the stellar halo, using \cite{horta23} selections for GSE and Thamnos \citep{Koppelman2019_thamnos}, and \cite{naidu2020} for the Sgr stream, Sequoia \citep{Myeong2019_seq} and LMS-1 \citep{Yuan2020_lms1}. The criteria for selecting Helmi streams \citep{Helmi99} members are the same in both works (from \citealt{Koppelman2019_helmi}).

\begin{figure}[t]
\centering
  \includegraphics[width=\columnwidth]{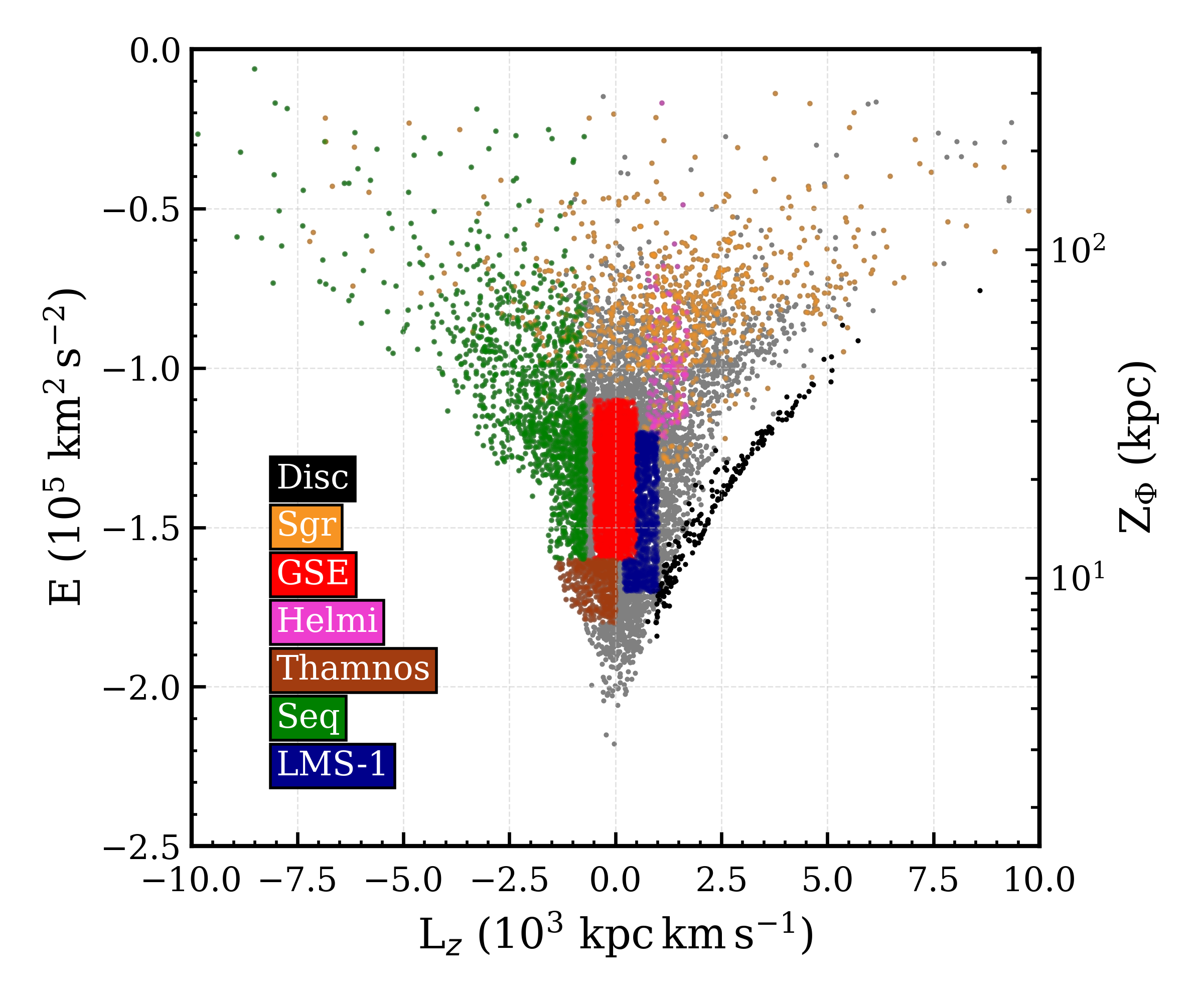}
  \caption{Lindblad diagram highlighting disc stars and the accreted structures identified: Sagittarius (Sgr), Gaia-Sausage/Enceladus (GSE), Helmi streams, Thamnos, Sequoia (Seq) and LMS-1. The median values from the MC simulation were adopted to represent each star. The auxiliary y-axis $\rm Z_\Phi$ indicates the corresponding Z position for a star with potential energy equals to the total energy E. }
  \label{fig:struc}
\end{figure}

The distribution of the selected structures in the Lindblad diagram is presented in Figure \ref{fig:struc} using the median values of energy and $\rm L_z$. 
$\rm Z_\Phi$ illustrates the extension of the structures, indicating that the vertical position of a star with potential energy equals to the orbital energy E. 
The selection of accreted structures was performed in each MC realization, resulting in a slightly different number of stars but a consistent fraction of the sample. The criteria used and the fraction of BHB stars found are presented in Table \ref{tab:estr}.

\begin{table}[h]
\centering
\caption{List of criteria used to select the accreted structures, the fraction of the sample comprised in each one and the references for the criteria adopted. The structures were identified based on: energy (E), angular momentum components ($\rm L_z$, $\rm L_y$, and $\rm L_{perp} = \sqrt{L_x^2 + L_y^2}$), and circularity ($\eta = \rm L_z/L_c$, being $\rm L_c$ the angular momentum for a circular orbit at the same energy).} 
\label{tab:estr}
\begin{adjustbox}{max width=\columnwidth}
\hspace{-1.3cm}
\begin{threeparttable}
  \begin{tabular}{llcc}
  \multicolumn{1}{l}{ } &
  \multicolumn{1}{c}{Criteria} &
  \multicolumn{1}{c}{Fraction} &
  \multicolumn{1}{c}{Ref.} \\
  \hline \hline 

  \multirow{1}{1.6cm}{Sgr} & $\rm L_y < 0.3 \cdot L_z - 2.5 \cdot 10^3$ & \multirow{1}{1cm}{0.09} & \multirow{1}{0.6cm}{1} \\ \hline

  \multirow{2}{1.6cm}{GSE} & $\rm |L_z| < 0.5 \cdot 10^3 \ kpc \, km \, s^{-1}$; & \multirow{2}{1cm}{0.33} & \multirow{2}{0.6cm}{2} \\ 
   & $\rm -1.6 < E \ (10^5 \ km^2 \, s^{-2}) < -1.1$ & \\ \hline
   
  \multirow{2}{1.6cm}{Helmi streams} & $\rm 0.75 < L_z \ (10^3 \rm \ kpc \, km \, s^{-1}) < 1.7$; & \multirow{2}{1cm}{0.01} & \multirow{2}{0.6cm}{3} \\
   & $\rm 1.6 < L_{perp} \ (10^3 \rm \ kpc \, km \, s^{-1}) < 3.2$ & \\ \hline
   
  \multirow{2}{1.6cm}{Thamnos} & $\rm -1.8 < E \ (10^5 \ \rm km^2 \, s^{-2}) < -1.6$; & \multirow{2}{1cm}{0.05} & \multirow{2}{0.6cm}{2} \\
   & $\rm L_z < 0 \ kpc \, km \, s^{-1}$ & \\ \hline
   
  \multirow{3}{1.6cm}{Sequoia} & $\rm E > -1.6 \cdot 10^5 \ \rm km^2 \, s^{-2}$; & \multirow{3}{1cm}{0.15} & \multirow{3}{0.6cm}{1} \\
   & $\rm L_z < -0.7 \cdot 10^3 \ kpc \, km \, s^{-1}$;  & \\ 
   & $\eta < -0.15$ & \\ \hline
   
  \multirow{2}{1.6cm}{LMS-1} & $\rm 0.2 < L_z \ (10^3 \ kpc \, km \, s^{-1}) < 1$; & \multirow{2}{1cm}{0.09} & \multirow{2}{0.6cm}{1} \\
   & $\rm -1.7 < E \ (10^5 \ km^2 \, s^{-2}) < -1.2$ & \\ 
  \hline \hline 
  \end{tabular}
  \begin{tablenotes}
      \small
      \item \hspace{0.8cm} (1) \cite{naidu2020}
      \item \hspace{0.8cm} (2) \cite{horta23}
      \item \hspace{0.8cm} (3) \cite{Koppelman2019_helmi}
      
    \end{tablenotes}
  \end{threeparttable}
  \end{adjustbox}
\end{table}

Our aim is not to analyze the structures individually, but to investigate the stars not comprised by them. 
Based on the parameters used for the selection and the absence of detailed chemical abundances, we cannot and do not intend to have the purest samples for the accreted structures. Therefore, we chose not to distinguish Sequoia, Arjuna and I'itoi, since they are differentiated only by the metallicity range \citep{naidu2020}. 

Furthermore, we also chose not to impose a metallicity cut for other structures. 
As discussed in \cite{naidu2020}, the dynamical selection of Thamnos and LMS-1 results in a bimodal metallicity distribution. \cite{naidu2020} defined a cut to separate the peaks to select the most metal-poor stars as members of the structures, arguing that the second peak may correspond to GSE stars. Here, we chose to keep the selection based only on orbital parameters, as removing possible GSE stars as part of another accreted structure is beneficial to the subsequent analysis.

After removing accreted structures, we obtain a sample that corresponds to $\sim 27\%$ of the complete sample in each realization. Hereafter, we refer to this as the clean sample and the analysis are done with the median values of the realizations. We recall that we do not attempt to achieve a ``clean'' sample in the strict meaning of the word, and only prominent structures were identified. 
Therefore, we may have stars belonging to those and other accretions in the clean sample. Most of the inner halo is known to be composed by the GSE (\citealt{Mackereth2019}, \citealt{naidu2020}, \citealt{Wu2022}), and the simple approach employed here will not remove it entirely. As shown by \cite{JeanBap17} and \cite{Pagnini22}, different structures may overlap in configuration spaces due to dynamical friction, preventing a simple selection. We also did not attempt to select members in stellar streams.

\begin{figure}[t]
\centering
  \includegraphics[width=\columnwidth]{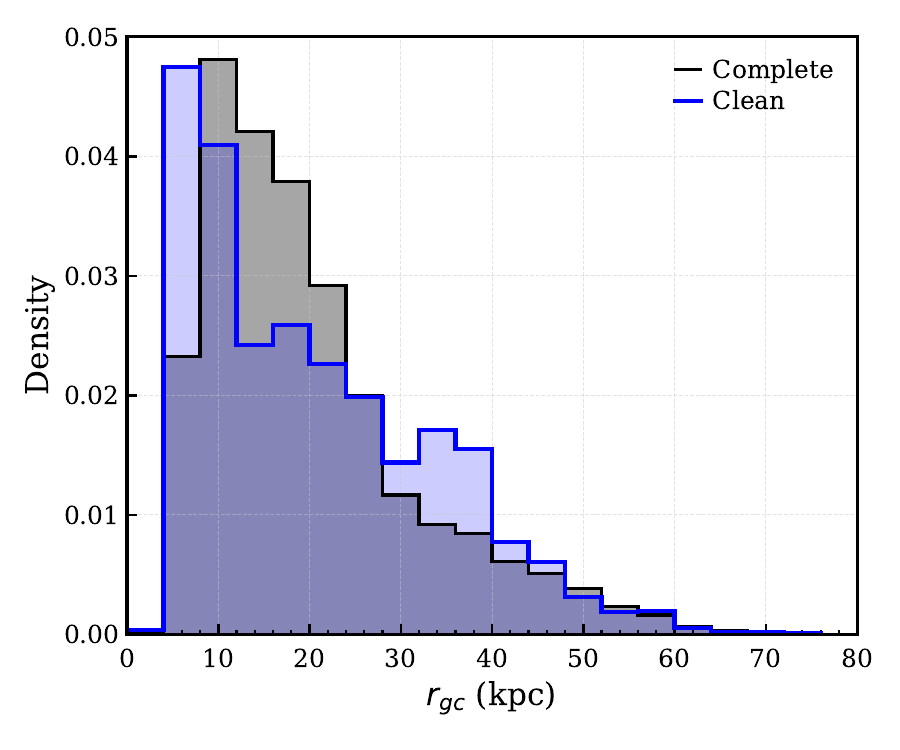}
  \caption{Density distribution of median Galactocentric distance for complete (grey) and clean (blue) samples.}
  \label{fig:r_dist}
\end{figure}

Figure \ref{fig:r_dist} shows the distribution of Galactocentric distance median values for the complete and for a clean sample when the selection is performed in median values, as in Figure \ref{fig:struc}. 
Many stars are removed near the GSE apocenter ($\sim 20$ kpc, \citealt{Deason2018}), shifting the distribution peak to around 6 kpc in the clean sample, and generating a second peak between 30 and 40 kpc.

\section{Kinematic anisotropy}
\label{sec:amostras}

Knowing the kinematics of the stellar halo is a fundamental step to estimate the dynamical mass of the MW. The traditional Jeans approach requires measurements of the velocity dispersions ($\sigma_i = \langle v_i^2 \rangle - \langle v_i \rangle^2$) in order to derive the mass inside a given radius. 
For simplicity, the Jeans equation can be arranged in a way that the velocity anisotropy parameter ($\beta$) can be defined from tangential ($\sigma_t$) and radial ($\sigma_r$) velocity dispersions as

\begin{equation}
    \beta \equiv 1 - \frac{\sigma_t^2}{\sigma_r^2}, 
\end{equation}

\noindent
as first presented in \cite{Binney1980}. This parameter describes the orbital structure of a system, measuring the dominance of radial and tangential velocity dispersion components. 
However, being an asymmetric function unlimited for negative values, it is not possible to directly compare radially- and tangentially-biased estimations.

Considering this, we opted for presenting the orbital structure of the halo using 

\begin{equation}
    \beta_s \equiv \frac{2\sigma_r^2 - \sigma_\theta^2 - \sigma_\phi^2}{2\sigma_r^2 + \sigma_\theta^2 + \sigma_\phi^2},
\end{equation} \label{eq:anis_index}

\noindent
which corresponds to the symmetrized anisotropy parameter presented in \cite{Read2006} for a spherical coordinate system. For conciseness, we refer to this symmetrized velocity anisotropy parameter simply as ``\textit{anisotropy index}'' ($\beta_s$). 

The index $\beta_s$ preserves the behavior observed with $\beta$ (see the comparison in Appendix \ref{ap:B}), with negative/positive values of $\beta_s$ indicating that the group of stars is dominated by the dispersion in tangential/radial velocity components, and $\beta_s = 0$ represents an isotropic sample. $\beta_s$ is defined in a symmetrical interval, limited in $[-1, 1]$, which reduces the distortion in the tangential behavior. 
Symmetrized anisotropy parameters have been used for numerical applications (\citealt{Read2017}, \citealt{Mamon2019}), as it enables the construction of tangentially-dominated models. 
However, in Galactic studies, the orbital structure of the halo has been described by the traditional anisotropy parameter $\beta$. $\beta$ and $\beta_s$ can be easily converted\footnote{The parameter $\beta$ can be obtained from $\beta_s$ using $\beta = \frac{2\beta_s}{1+\beta_s}$}, and a comparison between the values estimated here is presented in Appendix \ref{ap:B}.

\subsection{Anisotropy profiles}
\label{sub:beta}

In this work, the samples were divided in bins of 1 kpc from 6 to 10 kpc, 2 kpc between 10 and 28 kpc, and 3 kpc for stars until 40 kpc. We kept the first and last two bins larger to avoid having fewer stars. The bins were chosen in order to detail the profile, while maintaining $\gtrsim100$ stars per bin after removing the structures. The distance intervals and corresponding median values for velocity dispersions in both complete and clean sample are presented in Table \ref{tab:bins} in the Appendix \ref{ap:B}.

Figure \ref{fig:vel_disp} shows the profiles for the velocity and dispersion in each component for both samples. The median values of the MC simulation are represented in the plot. 
The top panels show mean values of velocity for each bin in both samples. 
The most striking differences between the samples appear for tangential components. 
There is an increase in the polar component for the clean sample after 20 kpc, going from close to zero to $36 \ \rm km \, s^{-1}$. As most retrograde stars fall in the cuts for accreted structures, the clean sample shows a prograde rotation, with $v_{\phi}$ between $37 \ \rm km \, s^{-1}$ and $68 \ \rm km \, s^{-1}$ for all Galactocentric radii. 

\begin{figure*}[t]
\centering
  \includegraphics[width=2\columnwidth]{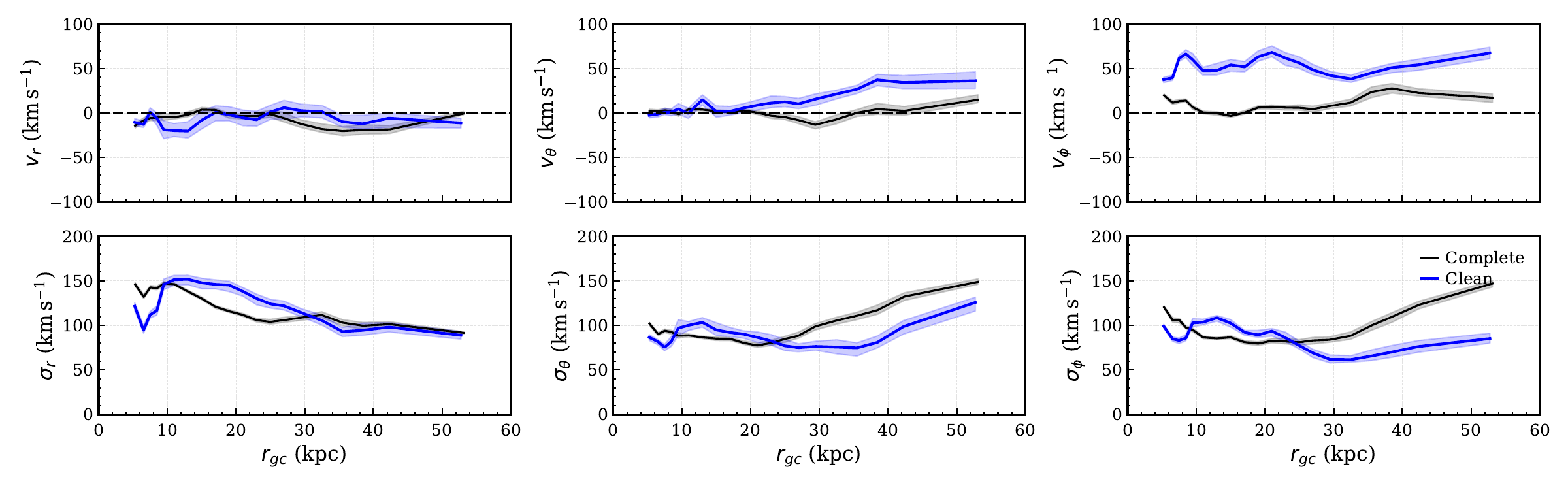}
  \caption{Velocities profiles (top row) and velocities dispersions (bottom row) for complete and clean samples. Median values were chosen to represent each distance bin. The bands correspond to the values obtained within the 16th and 84th percentiles from the Monte Carlo simulation.}
  \label{fig:vel_disp}
\end{figure*}

The bottom panels in Figure \ref{fig:vel_disp} present the velocity dispersion profiles. 
In the complete sample (black thin line), all velocity components show a pronounced transition.
To determine the distance at which the trends change, we used a segmented linear regression performed with \texttt{piecewise-segmented} Python package \citep{Pilgrim2021}, taking into account the Bayesian information criterion \citep[BIC;][]{bic} to define the number of breaks. The slopes and breaks measured are presented in Table \ref{tab:slop} in the Appendix \ref{ap:B}. 
We find a breakpoint at 23 kpc for $\sigma_r$, which marks a transition between a sharp and a shallow decline. On the other hand, for $\sigma_\theta$, the profile changes from a negative to a positive slope at 21 kpc. 
The tangential component $\sigma_\phi$ presents two breaks, at 11 and 29 kpc, defining a region where the profile remains nearly flat between an inner decline and an outer rise.  

The clean sample (blue thick line) shows more complex profiles. Different from the trends observed with the complete sample, the clean sample presents a positive slope after an inner drop inside $\sim 7$ kpc for all components. 
The profiles reach a peak at around 10 kpc, after which they decrease smoothly and reach a subsequent break between 31 and 36 kpc. Both tangential components present a clear positive outer trend, while $\sigma_r$ becomes nearly flat.

\begin{figure}[t]
\centering
  \includegraphics[width=\columnwidth]{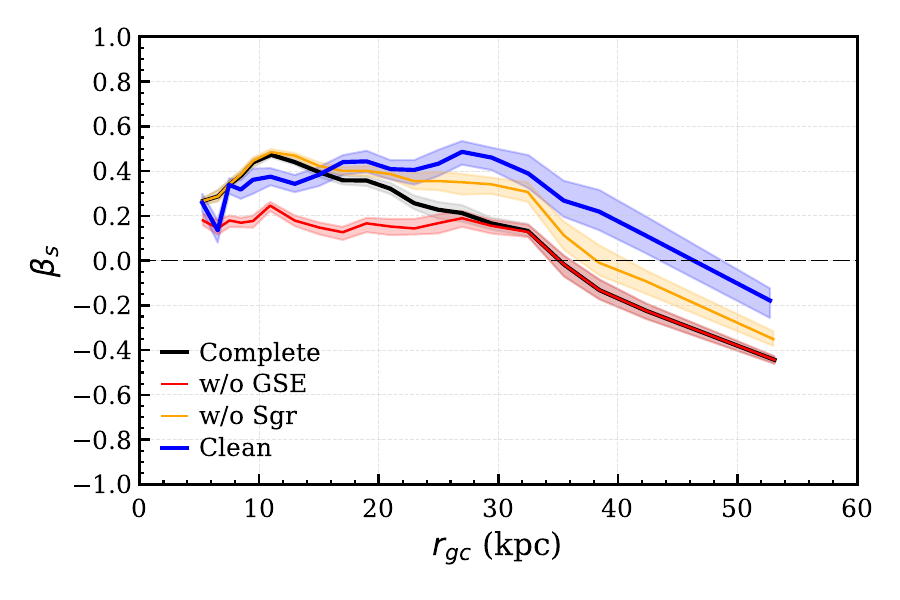}
  \caption{Anisotropy profile for all halo stars (black line) and removing only GSE (red line) or Sgr (yellow line) members, and for the sample after accreted structures are removed (blue line). The lines correspond to the median values and the respective bands indicate the limits using 16th and 84th percentiles obtained with MC simulation.}
  \label{fig:beta}
\end{figure}

Figure \ref{fig:beta} shows the $\beta_s$ profile obtained for the complete sample (black line), the sample removing stars associated to GSE (red line), and associated to Sgr (yellow line), and the clean sample (blue line). 
In the complete sample, we observe that radial dispersion dominates within 35 kpc. 
The segmented regression indicates four regimes. Within 11 kpc, $\beta_s$ grows rapidly, going from $\beta_s = 0.27$ to $\beta_s = 0.47$. Then, a steady decline begins, reaching $\beta_s = 0.13$ at 33 kpc. 
After this point, the profile falls sharply, becoming tangentially dominated beyond 37 kpc, reaching $\beta_s = -0.44$ at the edge of the observed distance range. 
The profile becomes less radial and approximately constant with $\beta_s \approx 0.2$ removing only GSE members. 
While GSE is restricted to the inner halo ($r_{gc} <$ 30 kpc), Sgr stream significantly affects the outer part, being the major responsible for the tangential behavior at 40 kpc. When the Sgr stream is removed, we note a sharper drop between 32 and 37 kpc (with a slope $a_3 = -0.064 \pm 0.005 \ \rm kpc^{-1}$) and a final smooth decrease compatible to observed for the complete sample ($a_4 = -0.024 \pm 0.001 \ \rm kpc^{-1}$). 
The clean sample presents some notable differences compared to the complete halo. 
With smaller samples and larger uncertainties, some oscillations appear. $\beta_s$ increases with the distance, with a breakpoint at 28 kpc, reaching $\beta_s = 0.49$. 
At this point, the profile drops, approaching isotropy beyond 45 kpc, with a minimum of $\beta_s = -0.18$.

\subsection{Metallicity/color effects on the anisotropy}
\label{sub:connections}

As a result of a mixture of structures in the halo, a correlation with metallicity appears. The initial works of \cite{Hattori2013} and \cite{Kafle2013} showed that BHB stars with $\rm[Fe/H] > -2$ exhibit higher anisotropy values inside 20 kpc, being radially dominated, while more metal-poor stars present more tangential orbits. 
More recently, \cite{Vickers2021} analyzed $\sim 2700$ BHB stars observed by LAMOST at $r_{gc} < 25$ kpc and also observed an increase in anisotropy with metallicity. Unlike previous works, they found an approximately constant radial profile for both groups. 
Using a clustering algorithm to identify and remove accreted structures, \cite{Bird2021} observed that the relation between anisotropy and metallicity persists, which could indicate that a significant part of accreted stars had already been virialized and compose the \textit{smooth halo}.
Here, we investigate this correlation with the kinematic selection. 

\begin{figure}[t]
\begin{center}
\begin{overpic}[width=\columnwidth]{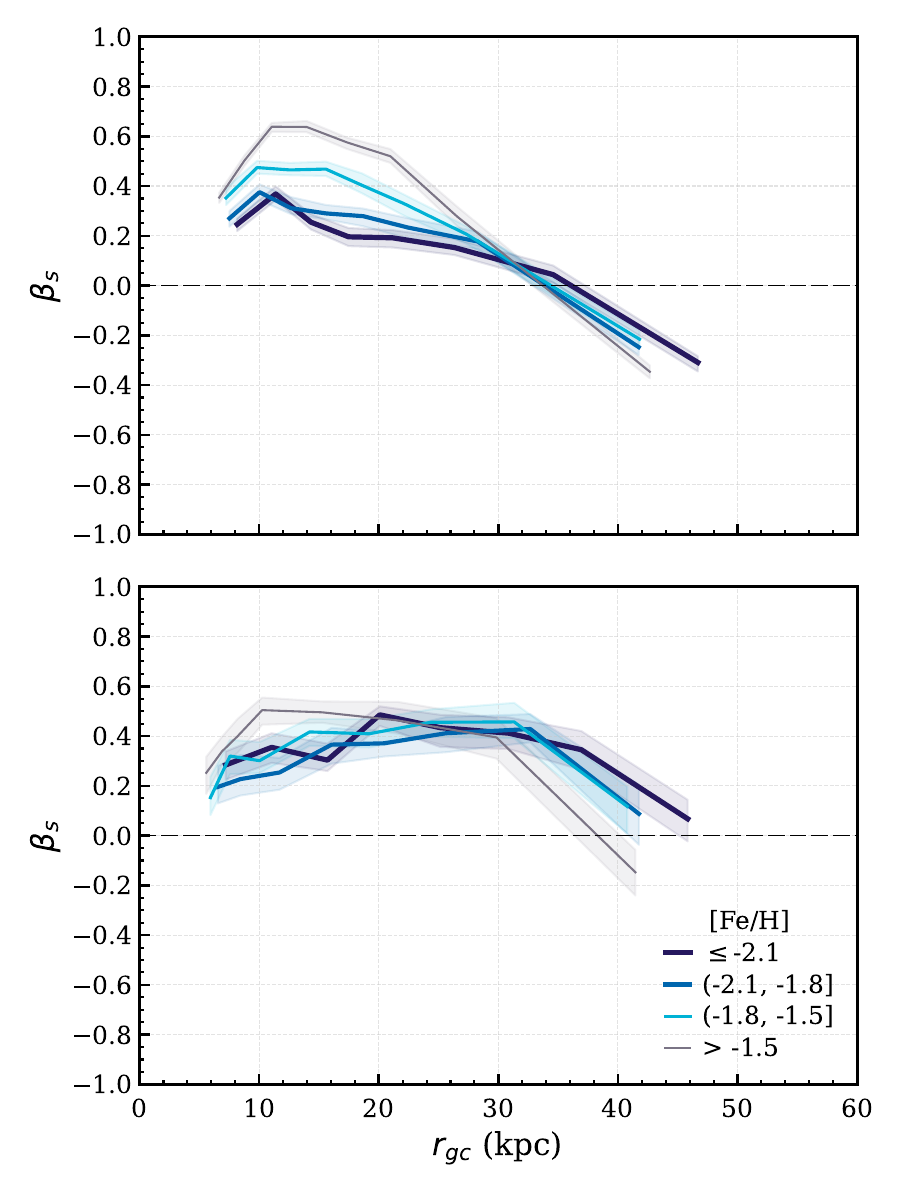}
    \put(14,58){\textbf{Complete sample}}
    \put(14,12){\textbf{Clean sample}}
\end{overpic}
\caption{Anisotropy profiles dividing the complete (top panel) and clean sample (bottom panel) in metallicity bins. Metallicity are indicated by color and thickness. The lines correspond to the median values and the respective bands indicate the limits using 16th and 84th percentiles obtained with MC simulation.}
\label{fig:beta_feh}
\end{center}
\end{figure}

\begin{figure}[t]
\begin{center}
\begin{overpic}[width=\columnwidth]{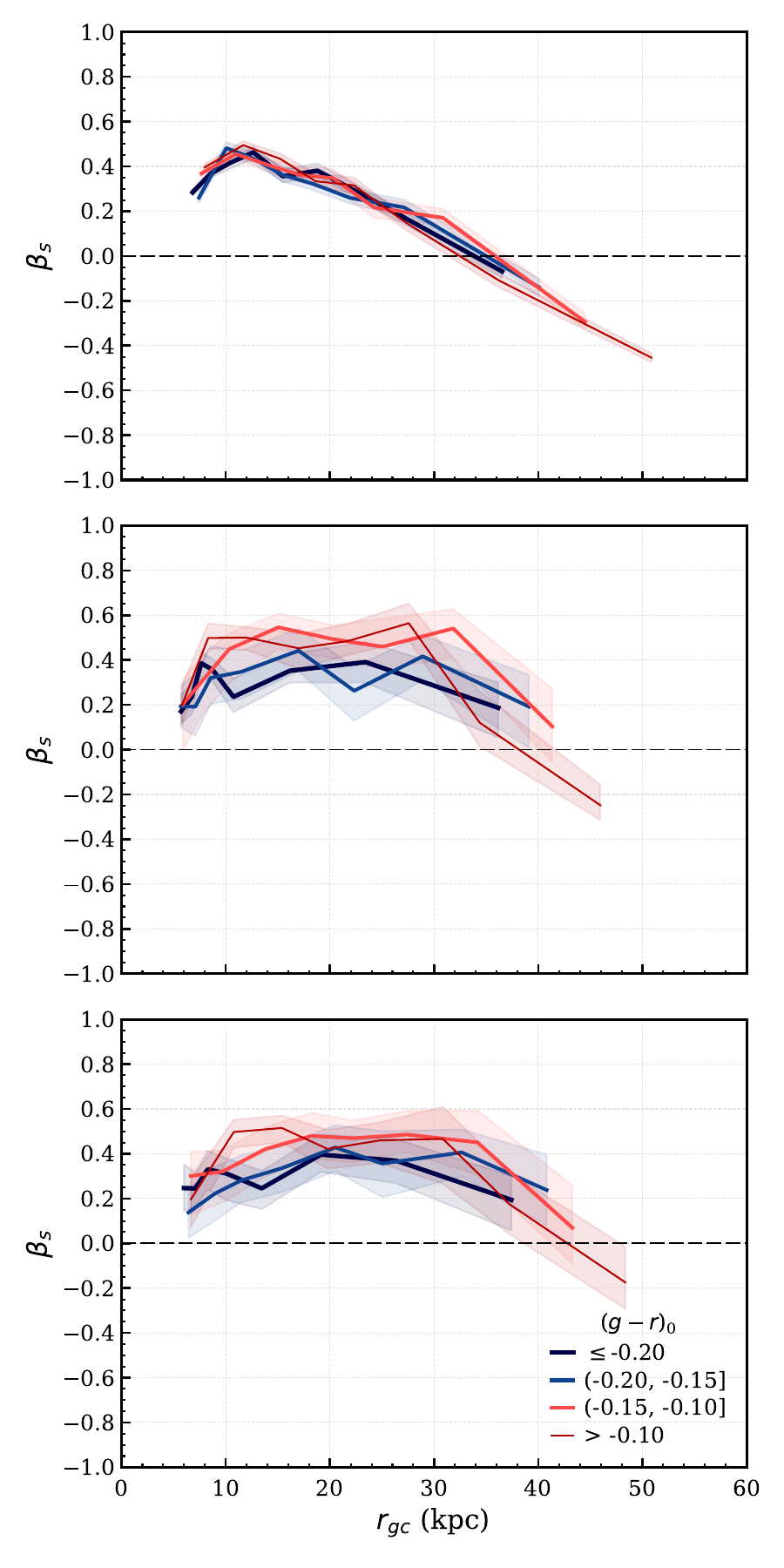}
    \put(10,71.8){\textbf{Complete sample}}
    \put(10,42){\textbf{Clean sample}}
    \put(10,40){\textbf{([Fe/H] $\boldsymbol{> -1.8}$)}}
    \put(10,10.3){\textbf{Clean sample}}
    \put(10,8.3){\textbf{([Fe/H] $\boldsymbol{< -1.8}$)}}
\end{overpic}
\caption{Anisotropy profiles for color bins dividing the complete (top panel) and clean sample ([Fe/H] $>-1.8$, middle panel; and [Fe/H] $< -1.8$, bottom panel). Color bins are indicated by color and thickness. The lines correspond to the median values and the respective bands indicate the limits using 16th and 84th percentiles obtained with MC simulation.}
\label{fig:beta_gr}
\end{center}
\end{figure}

Figure \ref{fig:beta_feh} presents the $\beta_s$ profile in bins of metallicity. The metallicity intervals and the bins of Galactocentric distance were defined to divide the sample into an approximated equal number of stars. The x-axis position indicates the median $r_{gc}$ in each bin. 
In the complete sample (top panel), the bins comprehend, approximately, 318 stars each, while, in the clean sample (bottom panel), we have $\sim 85$ stars each. 
The top panel shows that, as the metallicity increases, the $\beta_s$ profiles become more radially-biased inside 25 kpc for the complete sample, in agreement with previous findings. 
The most metal-poor profile (purple line) reaches a maximum of $\beta_s = 0.37$ at 11 kpc, while the most metal-rich (gray line), reaches $\beta_s = 0.64$ at the same distance. 

All $\beta_s$ profiles show a decreasing trend, being tangential in the last bin. We can also notice the inversion of the correlation at larger distances, with the more metal-rich subset becoming the more tangentially dominated. 
The clean sample in the bottom panel shows that the relationship between $\beta_s$ and [Fe/H] is almost erased after removing the accreted structures. 
The lack of correlation between metallicity and $\beta_s$ in our clean sample reinforces that this relation is driven by merger events. 
We remind the reader that $\beta_s$ preserves the behaviour observed using $\beta$.

If the correlation between kinematic and metallicity is a signature of the accretion history, we would expect a correlation also with the age of the stars. 
With BHB stars, we can infer relative ages from photometric information. Keeping the metallicity constant, a difference in the color of the BHB stars reflects a difference in their ages, with bluer BHB stars being older than redder BHB stars.

To investigate whether this correlation can be detected, we present an exploratory analysis of correlation between $\beta_s$ and color in Figure \ref{fig:beta_gr}. The samples were divided into bins of 0.05 mag in $(g-r)_0$ and the clean sample was divided at $\rm [Fe/H] = -1.8$ to reduce the effect of the metallicity. 
With the complete sample (top panel), the $\beta_s$ profiles possess a similar trend, showing no relation to the color of the stars. 
A small distinction, which cannot be accounted for the uncertainties, is noted in the first bin, where redder stars are more radially-biased ($\beta_s \sim 0.40$) than the blue ones ($\beta_s \sim 0.28$), with bluer stars concentrating closer to the center of the galaxy, as can be noticed by the difference in the x-axis position. 
The higher $\beta_s$ observed for redder stars is mainly the result of a larger velocity dispersion in the radial component, with an increase of $34 \rm \ km\,s^{-1}$ for reddest profile compared to the bluest one. Also, the tangential components inside 20 kpc shows that blue BHB stars are dynamically colder than red BHB stars.

On the other hand, when analyzing the clean sample (middle and bottom panels), redder ($(g-r)_0 > -0.15$) and bluer stars ($(g-r)_0 < -0.15$) draw distinct profiles. Between 10 and 20 kpc, independent of the metallicity range, redder stars are more dominated by radial dispersion compared to bluer stars. 
The correlation between $\beta_s$ and color observed with the clean sample is not solely a consequence of the relation between metallicity and color for BHB stars, as can be observed with the metallicity separation between the middle and bottom panels. 
Therefore, it may reflect differences in the stellar ages.

\section{Discussion}
\label{sec:disc}

In this work, we explored the anisotropy of the MW halo with BHB stars out to 70 kpc. Our sample allows us to refine anisotropy profiles found by preceding works, checking for the presence of features, such as dips and peaks. Here, we combined spectroscopic and astrometric data to investigate how the most massive accretion events currently known affect the anisotropy of the stellar halo.

As discussed in Section \ref{sec:strucs}, our selection is mainly based on kinematic and dynamic parameters, which will bias the velocity distributions. Most retrograde stars are excluded, generating a prograde halo. Still, this simple approach lead to some interesting results.

\subsection{Anisotropy profiles in the literature}
\label{sub:comp_lit}

Previous works relying on distribution functions to describe the anisotropy found predominantly tangential halo beyond 15 kpc (\citealt{Kafle2012, Kafle2013}, \citealt{Hattori2013}). However, \cite{Hattori2017} showed that using only line-of-sight velocities can underestimate the anisotropy for distances greater than $r_{gc} = 15$ kpc. Since Gaia DR2, some works have been performed analyzing full phase space. However, their samples hardly reach distances as far as 50 kpc, resulting in inconsistent anisotropy values. 

Analyzing $\sim 3000$ BHB stars with SDSS/SEGUE and Gaia DR2 data, \cite{Lancaster2018} derived the anisotropy of halo stars within four radial bins, excluding members of the Sgr stream. They found that $\beta$ declines from 20 to 40 kpc, reaching an isotropic to slightly tangential behavior in the outermost bin. Converting $\beta_s$ to $\beta$, we see that our profile is in good agreement with their results, showing $\beta \sim 0.6$ at 15 kpc, but a slightly negative value at 40 kpc (see the corresponding values of $\beta$ in Appendix \ref{ap:B}). 
Our profile removing GSE also agrees with their isotropic fit within 30 kpc, with $0.2 < \beta < 0.4$. 

Later, \cite{Bird2021} showed the anisotropy profile with $\sim 4000$ BHB and more than 10000 K giants stars out to 60 kpc using SDSS/SEGUE and LAMOST data, complementing a previous work \citep{Bird2019}. Up to 25 kpc, they found an almost constant velocity anisotropy profile for BHB stars, oscillating between $\beta \sim 0.5 - 0.7$, and a more radial profile for K giants, increasing from $\beta \sim0.5$ to 0.8. Both stellar types show a decline beyond 20 kpc for the complete samples, in agreement with the findings presented in this work. On the other hand, no tangential anisotropy is observed at any distance, in contrast with our results, where the complete halo becomes tangentially dominated beyond 35 kpc. 
Despite the difference in the binning, we believe this disagreement could be due to the different sample size. Beyond 40 kpc, our sample comprises more than 800 stars, better mapping the kinematics in the outer regions.

\cite{Wu2022} also observed a decrease in $\beta$ for both BHB and K giant stars around 40 kpc, with BHB stars reaching negative values. Analyzing RRL stars, \cite{Liu2022} identified the previous transition, with a high anisotropy inside 20 kpc ($\beta \sim 0.8$), decreasing to a lower value ($\beta \sim 0.5$). 
Simulations indicate that the hierarchical halos are radially-biased, with $\beta$ increasing with distance in the inner region, and negative values are probably tracing unrelaxed accreted structures, such as Sgr and other streams (e.g., \citealt{Rashkov2013},  \citealt{Mondal2024}). 
Removing Sgr streams here does not erase the tangential signature in the outer halo, possibly pointing to the influence of other streams.

Recently, \cite{Li2026} analyzed a large sample of K giant stars observed by DESI ($\sim 60,000$) and reported a similar velocity anisotropy profile, with a peak close to 10 kpc followed by a steady decrease, reaching isotropy around 85 kpc. 
Our trend is consistent with their findings, with a peak at 10.7 kpc (see Table \ref{tab:slop}), and the difference in the point of isotropy (35 kpc with BHB stars) would be expected since K giants present higher anisotropies at all distances. Hence, the transition to tangential values would appear at a smaller distance for BHB stars.

\subsection{Influence of accretion events}
\label{sub:acrecoes}

MW accretions left many imprints that contribute to the anisotropy profile. Works using the HALO7D survey (\citealt{Cunningham2016}, \citealt{Cunningham2019}, \citealt{McKinnon2023}) found that the anisotropy changes for different fields. 
Selecting the structures individually, we inspect their impact in the anisotropy profile. We note that Sgr stream contributes to the halo being less radial beyond 20 kpc with a difference of up to $\Delta\beta_s = 0.17$ (Figure \ref{fig:beta}). When this structure is removed, the last bin remains tangentially dominated, indicating that it is not the only unrelaxed accretion to be removed. On the other hand, GSE strongly affects the inner region, and its exclusion leads to an approximately constant anisotropy profile, with $\beta_s \approx0.2$ for $r_{gc} < 30$ kpc. However,  the GSE selection includes more metal-poor stars ([Fe/H] $< -2$), which may pertain to other accreted systems or to the in-situ component, increasing the anisotropy observed. 
As other non-radial structures are removed, these values rise again, suggesting that the ancient massive merger cannot entirely explain a radially-dominated stellar halo \citep{Deason2024}.

\cite{Amorisco2017} and \cite{Vasiliev2022} presented N-body simulations of mergers and showed that radialization can be a natural consequence of dynamic interactions for massive accretions. During the disruption, the asymmetry generated in the satellite induces self-friction, which leads to a decrease in angular momentum. This would explain why high $\beta$ (and $\beta_s$) values dominate a large portion of the halo. According to the experiments performed by \cite{Vasiliev2022}, unless the system is of low mass (mass ratio $<$ 1:20) and initially presents a more circular orbit ($\eta > 0.5$), the torque generated induces the radialization of the orbits. Therefore, the presence of a significant, though less massive than a GSE, merger would lead to a highly anisotropic halo, regardless of its initial orbit. The drop in $\beta_s$ profile observed in the outermost parts of the clean halo could be indicative of the end of the dominance of the more significant mergers and of the presence of streams, which are more resistant to this process and were not considered in this work. 

Another clear characteristic in the profile is the rise in anisotropy in the innermost region, both with and without accreted structures, with variation in the latter being more gradual. This is in agreement with simulations, where accreted halos show a radial anisotropy profile, decreasing from the outskirts to the center (e.g., \citealt{Rashkov2013}, \citealt{Wojtak2013}, \citealt{Wang2015}, \citealt{He2024}, \citealt{Mondal2024}). 
The increase in the profile can be seen in \cite{Kafle2013} and \cite{Bird2021} with K giants profiles, with a difference of $\Delta\beta \lesssim 0.2$. It is also noted by \cite{Li2026} and associated to thick disk contamination. With a more detailed binning and a more restrictive selection of disk stars, we can confirm the rising in the anisotropy profile. In both samples analyzed here, we observe a smaller $\beta$ for the innermost bins, with an increase of $\Delta\beta = 0.25$ until 11 kpc for the complete sample and a similar, but slower, increment for the clean sample of $\Delta\beta = 0.26$ inside 28 kpc. 

Using IllustrisTNG simulations, \cite{He2024} showed that an increase of $\beta$ with Galactocentric distance out to the scale radius is observed for accreted halo stars in galaxies with $10^{12.6} \textendash 10^{15} \ \rm M_\odot$. More massive galaxies display lower values inside the scale radius, being as low as 0.3 close to the center. 
\cite{He2024} propose as a possible explanation the exchange of angular momentum between accreted and in-situ particles, leading to circularization of the orbits. Accreted stars found closer to the center of the MW would originate from systems accreted in earlier times \citep{Amorisco2017}, and therefore had more time to undergo interactions. In that work, this effect is observed with both baryonic and dark matter. Confirming it in the MW stellar halo would indicate that this phenomenon might also be taking place in the dark matter halo.

The transition from a rising to decreasing behavior in the anisotropy profile could represent a transition between an in-situ dominated halo to an accreted-dominated halo. 
Analyzing red giant stars inside 10 kpc of heliocentric distance, \cite{Davies2025} found that an in-situ component dominates for $r_{gc} < 9$ kpc, consistent with cosmological simulations, and matching the transition observed in this work.

\subsection{Dips in the profiles}
\label{sub:dip}

Previous studies have pointed to the presence of a dip in the anisotropy profile around 20 kpc (\citealt{Kafle2012}, \citealt{King2015}, \citealt{Cunningham2016}, \citealt{Liu2022}). 
\cite{Loebman2018} investigated possible causes for a dip in the anisotropy profile analysing N-body+SPH and accretion-only simulations. They claimed that dips could be the result of a minor accretion, the passage of a satellite or a major merger, with each process imprinting different characteristics (depth and broadening). Therefore, identifying these features can provide information about the process that happened during the evolution of the MW. 

Using the dip definition in \cite{Loebman2018}, as a decrease $\ge 0.2$ in $\beta$ compared to the surroundings, no dip is clearly distinguishable in Figure \ref{fig:beta}. However, it does appear in the clean sample for stars with [Fe/H] $> -1.8$ and $-0.20 \leq (g-r)_0 < -0.15$ (Figure \ref{fig:beta_gr}). 
Comparing the velocity dispersions, we note that the dip is mainly caused by a decrease in the dispersion of the radial and polar components. It is also consistent with the mean apocenter for GSE stars measured with BHB stars in \cite{Deason2018}: $20\pm7$ kpc, which could indicate that the higher values observed between 10--20 kpc might be due to GSE members not covered by the criteria used.

\subsection{Old vs. young halo stars}
\label{sub:insitu_halo}

The connection between color and age for horizontal branch stars is well known (\citealt{Lee1994}, \citealt{Dotter2010}). Accounting for the correlation with metallicity (bluer, poorer), color can be used as a proxy for age.  
In the MW, we observe this in the stellar halo, where the color gradient exposes the age variation generated by the hierarchical formation (\citealt{preston1991}, \citealt{santucci_map}, \citealt{carollo2016}, \citealt{Das2016}). High-mass, old, accreted systems are found deep in the MW potential, while recent low-mass accretions populate regions far from the center (\citealt{Amorisco2017}, \citealt{Tau2025}). 
Looking at the bottom panels in Figures \ref{fig:beta_feh} and \ref{fig:beta_gr} we see a correlation with color that cannot be justified only by the metallicity.  

Moreover, a few noticeable distinctions can be observed in the bottom panel of Figure \ref{fig:beta_gr}. First, between 8 and 28 kpc, the metal-rich ([Fe/H] $> -1.8$) profile with $(g-r)_0 > -0.10$ display higher $\beta_s$ values compared to the one with $(g-r)_0 \leq -0.20$. The same result is observed for metal-poor ones, however, with an agreement between the profiles due to the higher uncertainties. 
Although we attempt to remove most prominent accreted structures in the sample, we recall that accreted stars are certainly present in the clean sample, given the difficulty of selecting all and only accretion events. With this in mind, the different anisotropies could reflect the different fraction of in-situ/accreted stars, with redder (younger) stars coming from accretion events. 

Second, the reddest profile presents a drop quickly beyond 30 kpc without correspondence in the other profiles. 
A possible interpretation could be that the lower $\beta_s$ values are due to stars in the outer part of the halo in more circular orbits. However, these stars present a median eccentricity around 0.7, consistent with the last bin in the light blue profile. Indeed, the dominance of the tangential component is caused by a larger dispersion in the polar velocity component, which could be due to the presence of stellar streams.

Studies conducted with different tracers have reported distinct anisotropies for the same radii. 
\cite{Bird2021} showed that K giant stars are  more radially dominated than BHB stars ($\Delta\beta \sim0.2$). 
Analyzing $\sim 3000$ RRL stars, \cite{Liu2022} noted a trend similar to BHB stars in \cite{Lancaster2018}. $\beta$ is highly radial ($\beta \sim 0.8$) out to 20 kpc, with a steady decline to the outer parts. The trend in the outer region agrees, with a decrease in the anisotropy observed in this work, but more modest for RRL ($\Delta\beta \sim 0.3$ vs. $\sim 1.1$ for BHB stars). 

As for K giants, RRL stars trace a higher anisotropy profile compared to BHB stars. \cite{Liu2022} claim that this is related to the difference in the metallicity. However, there are still differences in the estimated $\beta$ when comparing the profiles in a similar metallicity range. 
Contrary to the observed here, their more metal-rich subset ($\rm [Fe/H] > -1.7$) does not present an abrupt drop in the outer halo. Our BHB stars with $-2.1 < \rm [Fe/H] < -1.8$ do not surpass $\beta = 0.6$ as in \cite{Liu2022} for $-2.0 < \rm [Fe/H] < -1.7$. 
This is also true both for the complete sample and when only Sgr stream and GSE are removed.

Here, dividing the sample of BHB stars by color, we note differences in anisotropy that resembles the difference with stellar types, with younger BHB stars being more radially biased ($\beta_s \approx 0.44$) compared to older BHB stars ($\beta_s \approx 0.30$). 
This could point to the influence of dynamical mechanisms circularizing the orbits, making older stellar populations become more isotropic, as suggested by \cite{He2024} or to a different origin. To check if both subsets originate from a mixture of structures and become kinematically distinct due to dynamical interactions, or if one comes from an in-situ component originally more isotropic, requires chemical information for these stars.

\section{Summary \& Conclusions}
\label{sec:conc}

This work presents the velocity dispersion profiles for a sample of 10,323 BHB stars, combining spectroscopic observations from SEGUE, DESI, and LAMOST with Gaia DR3 astrometric data, that reaches the MW's outer halo at 70 kpc. Using photometric distances, line-of-sight velocities and proper motions, we obtain a detailed radial profile for the velocity anisotropy, and measure the impact of accreted structures selecting them by their orbital properties.

We observe a transition in the kinematic properties of the complete sample. The anisotropy profile switches from being dominated by radial velocities to dominated by tangential velocities at $r_{gc} = 35 \ \rm kpc$. To verify the effect of accreted stars in the profile, the major accreted systems were removed, leaving a sample that exhibits a more radially-biased anisotropy profile. 
These results would indicate that, even removing most of the accreted stars, no isotropic component could be detected. If an in-situ spherical halo existed in the beginning of the MW history, the anisotropy observed could imply that secular effects are responsible for the flattening observed in the present.

We also observe the rising of the anisotropy in the inner region for $r_{gc} < 11$ kpc, as observed with simulations, which might appear due to the stars undergoing tidal effects in the MW potential and losing memory of their origin. 
Verifying the predictions from simulations, such as the circularization in the inner region, is an important test to current galactic models, and also adds clues to understand the interaction between baryonic and dark matter.

We show that the correlation between anisotropy and metallicity is the result of accreted systems with different characteristics. We also present a first exploration of correlations between anisotropy and age using color cuts. 
After removing accretions, a correlation with color $(g-r)_0$ is observed, with redder stars presenting higher values of anisotropy. This dependence could indicate a correlation with age, as bluer BHB stars tend to be older than the redder ones with the same metallicity. Older stars being less radial also shows the effect of dynamical interactions closer to the center of the MW.

BHB stars are excellent objects to investigate the MW halo, with their photometric distances dismissing parallaxes and being more precisely determined than for K giants and with a lower observational cost than RRL stars. 
However, current works using BHB stars are limited by the absence of chemical abundances, particularly [$\alpha$/Fe]. Available spectroscopic catalogues providing [$\alpha$/Fe] have not being validated to stars with $\rm T_{eff} > 6500 \ K$. The high values observed with catalogues used in this work ([$\alpha$/Fe] $\gtrsim +0.6$) indicate an issue with these estimates to hot stars. 
Future spectroscopic surveys focusing on BHB stars could enrich Galactic evolution studies, with detailed chemical abundances improving the selection of members from accreted systems, while rotational velocities could help the identification of BHB stars, distinguishing it from dwarf stars \citep{kinman2000}.

\begin{acknowledgments}

This study was financed, in part, by the São Paulo Research Foundation (FAPESP), Brasil. Process Number \#2020/15245-2, \#2022/16502-4, \#2024/16510-2 and \#2026/01108-0). F.O.B and A.P.-V. acknowledge the DGAPA–PAPIIT grant IN112526. L.A. acknowledges support from the DGAPA–PAPIIT grant IG-101723 and IN110126. 
S.R. also acknowledgs the partial support from CNPq (Proc. 303816/2022-8) and CAPES. 
R.M.S. acknowledges CNPq (Proc. 307726/2025-8) for the financial support of this research. 
F.O.B. and A.P.-V. thank O. Valenzuela and A. Byström for the advices about the DESI sample.

This research has made use of the VizieR catalogue access tool, CDS, Strasbourg, France (\url{https://cds.u-strasbg.fr}). The original description of the VizieR service was published in \citet{vizier}.


This work has made use of data from the European Space Agency (ESA) mission Gaia (\url{https://www.cosmos.esa.int/gaia}), processed by the Gaia Data Processing and Analysis Consortium (DPAC, \url{https://www.cosmos.esa.int/web/gaia/dpac/consortium}). Funding for the DPAC has been provided by national institutions, in particular the institutions participating in the Gaia Multilateral Agreement.

Funding for the Sloan Digital Sky Survey IV has been provided by the Alfred P. Sloan Foundation, the U.S. Department of Energy Office of Science, and the Participating Institutions. SDSS acknowledges support and resources from the Center for High-Performance Computing at the University of Utah. The SDSS web site is www.sdss.org.
SDSS is managed by the Astrophysical Research Consortium for the Participating Institutions of the SDSS Collaboration including the Brazilian Participation Group, the Carnegie Institution for Science, Carnegie Mellon University, Center for Astrophysics | Harvard \& Smithsonian (CfA), the Chilean Participation Group, the French Participation Group, Instituto de Astrofísica de Canarias, The Johns Hopkins University, Kavli Institute for the Physics and Mathematics of the Universe (IPMU) / University of Tokyo, the Korean Participation Group, Lawrence Berkeley National Laboratory, Leibniz Institut für Astrophysik Potsdam (AIP), Max-Planck-Institut für Astronomie (MPIA Heidelberg), Max-Planck-Institut für Astrophysik (MPA Garching), Max-Planck-Institut für Extraterrestrische Physik (MPE), National Astronomical Observatories of China, New Mexico State University, New York University, University of Notre Dame, Observatório Nacional / MCTI, The Ohio State University, Pennsylvania State University, Shanghai Astronomical Observatory, United Kingdom Participation Group, Universidad Nacional Autónoma de México, University of Arizona, University of Colorado Boulder, University of Oxford, University of Portsmouth, University of Utah, University of Virginia, University of Washington, University of Wisconsin, Vanderbilt University, and Yale University.

Guoshoujing Telescope (the Large Sky Area Multi-Object Fiber Spectroscopic Telescope LAMOST) is a National Major Scientific Project built by the Chinese Academy of Sciences. Funding for the project has been provided by the National Development and Reform Commission. LAMOST is operated and managed by the National Astronomical Observatories, Chinese Academy of Sciences.

This research used data obtained with the Dark Energy Spectroscopic Instrument (DESI). DESI construction and operations is managed by the Lawrence Berkeley National Laboratory. This material is based upon work supported by the U.S. Department of Energy, Office of Science, Office of High-Energy Physics, under Contract No. DE–AC02–05CH11231, and by the National Energy Research Scientific Computing Center, a DOE Office of Science User Facility under the same contract. Additional support for DESI was provided by the U.S. National Science Foundation (NSF), Division of Astronomical Sciences under Contract No. AST-0950945 to the NSF’s National Optical-Infrared Astronomy Research Laboratory; the Science and Technology Facilities Council of the United Kingdom; the Gordon and Betty Moore Foundation; the Heising-Simons Foundation; the French Alternative Energies and Atomic Energy Commission (CEA); the National Council of Humanities, Science and Technology of Mexico (CONAHCYT); the Ministry of Science and Innovation of Spain (MICINN), and by the DESI Member Institutions: www.desi.lbl.gov/collaborating-institutions. The DESI collaboration is honored to be permitted to conduct scientific research on I’oligam Du’ag (Kitt Peak), a mountain with particular significance to the Tohono O’odham Nation. Any opinions, findings, and conclusions or recommendations expressed in this material are those of the author(s) and do not necessarily reflect the views of the U.S. National Science Foundation, the U.S. Department of Energy, or any of the listed funding agencies.

\end{acknowledgments}

%
\facilities{Gaia, Sloan, DESI, LAMOST}

\software{VizieR \citep{vizier}, AGAMA \citep{AGAMA}, numpy \citep{numpy}, matplotlib \citep{matplotlib}, astropy (\citealt{astropy2013, astropy2018, astropy2022})
          }


\appendix

\section{Offset adjustments between the catalogues}\label{ap:A}

\begin{figure}[t]
\centering
  \includegraphics[width=\columnwidth]{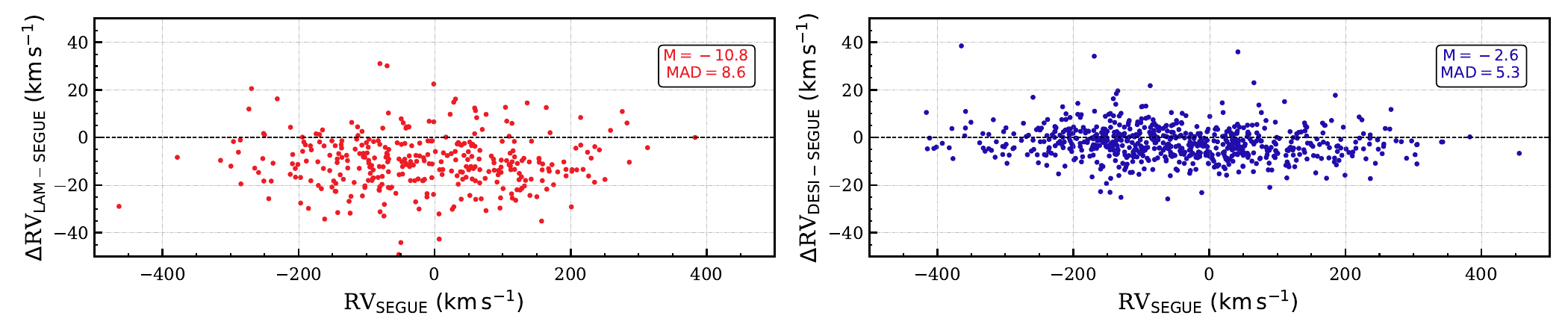}
  \includegraphics[width=\columnwidth]{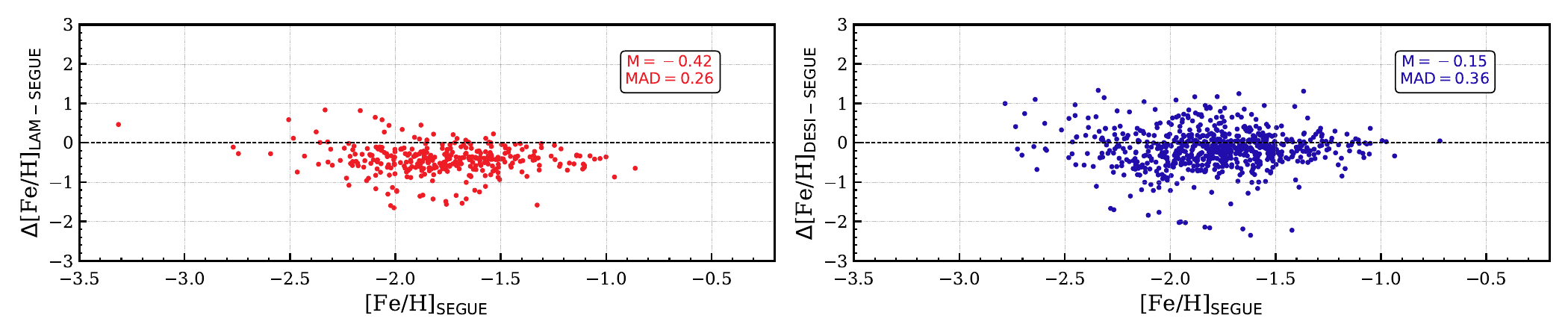}
  \caption{Difference between line-of-sight velocity (top panels) and metallicity (bottom panels) from LAMOST (right) and DESI (left) from SEGUE estimates. Median (M) and median absolute deviation (MAD) for each distribution is shown in the top right corner.}
  \label{fig:rv_corr}
\end{figure}

Line-of-sight velocities and metallicities for DESI and LAMOST samples were adjusted to match data from SEGUE to combine all three spectroscopic catalogues. 
Figure \ref{fig:rv_corr} shows how spectroscopic data for LAMOST (left) and DESI (right) differs from SEGUE data. We used common stars to evaluate the offset, finding 688 stars observed by DESI and 339 stars by LAMOST. 
We find a small offset for all parameters, indicated by the median (M) values in each plot. DESI and LAMOST measurements were corrected taking into account the medians, and median absolute deviations (MAD) were added in quadrature to the uncertainties.

\begin{figure}[t]
\centering
  \includegraphics[width=\columnwidth]{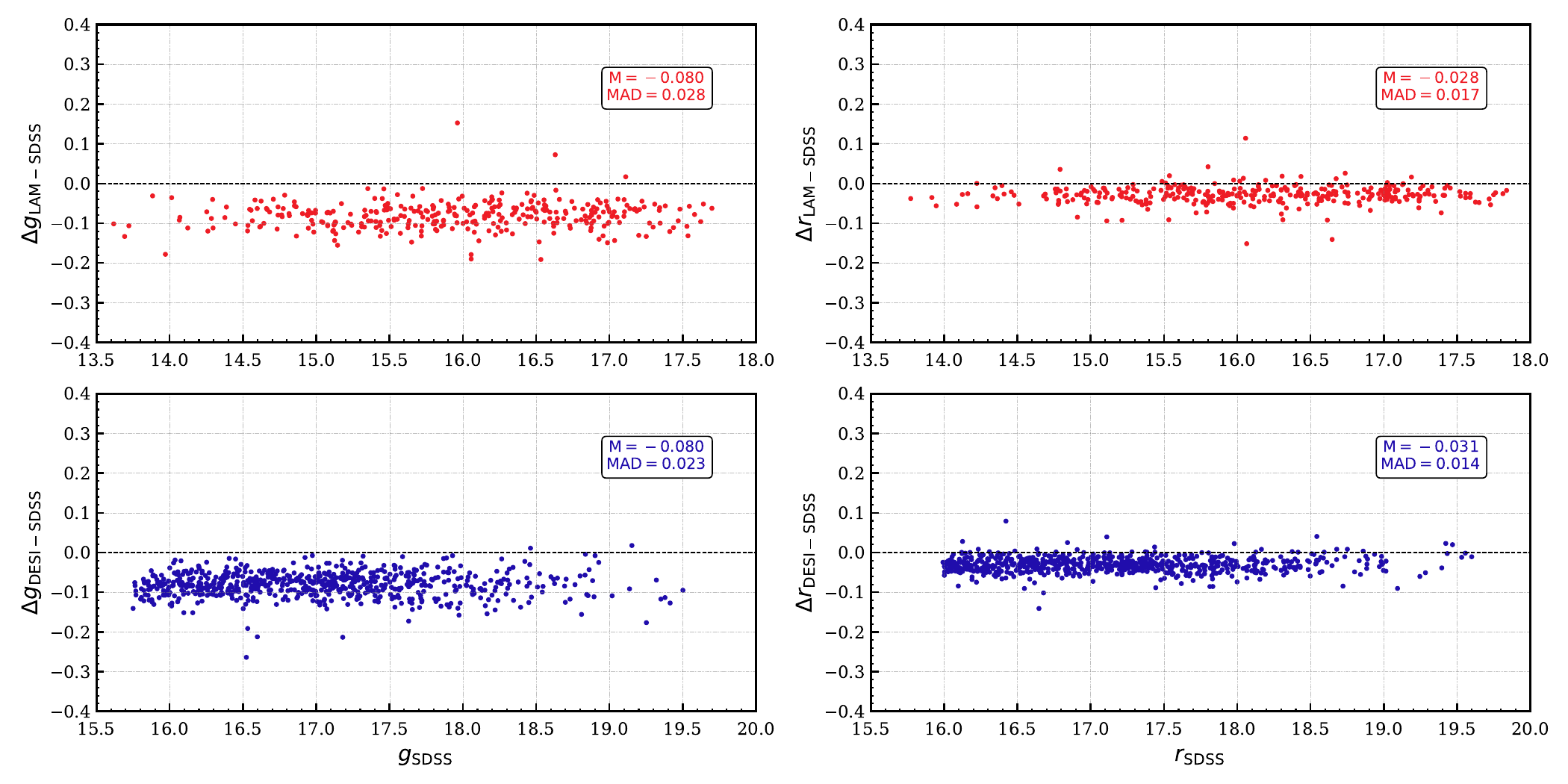}
  \caption{Difference between $g$-band (left panels) and $r$-band (right panels) from LAMOST (top panels) and DESI (bottom panels) from SDSS DR16 observations. Median (M) and median absolute deviation (MAD) for each distribution is shown in the top right corner.}
  \label{fig:phot_corr}
\end{figure}

Photometric distances were obtained using the absolute--magnitude estimation from \cite{barbosa22} with color $(g-r)_0$. For stars observed by DESI or LAMOST, SDSS $g$ and $r$ photometry was obtained from Gaia photometric system using the transformation provided by \cite{gaia_phot}. Figure \ref{fig:phot_corr} presents the comparison between transformed and observed SDSS bands for common stars in LAMOST (top panels) and DESI (bottom panels). As with the spectroscopic data, the transformed photometric values and uncertainties were adjusted using M and MAD.

\section{$\beta$ vs. $\beta_s$}\label{ap:B}

\begin{figure*}[!htb]
\begin{center}
\begin{overpic}[width=\columnwidth]{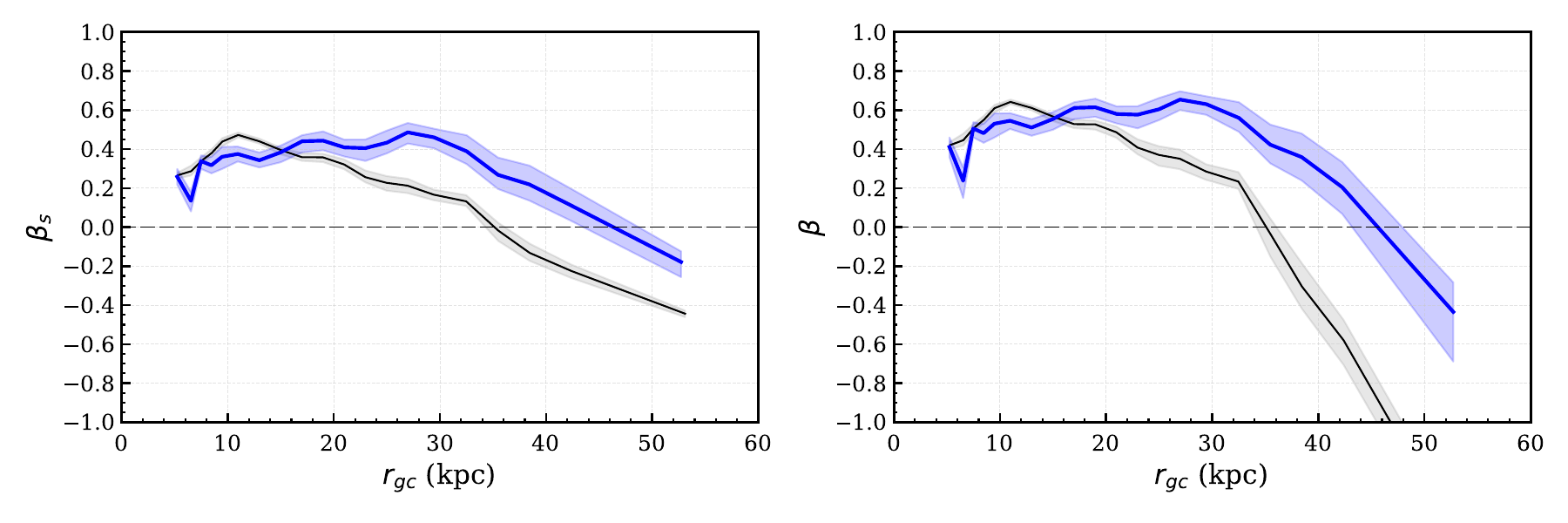}
\end{overpic}
\caption{Comparison between the kinematic anisotropy described by the anisotropy index ($\beta_s$, left panel) and the anisotropy parameter ($\beta$, right panel) for the complete and clean samples (black and blue lines, respectively).}
\label{fig:comp_beta}
\end{center}
\end{figure*}

Figure \ref{fig:comp_beta} shows the difference between the anisotropy parameter ($\beta$) and the anisotropy index ($\beta_s$) analyzed throughout this work. The profiles for the complete and clean samples are presented in both panels with black and blue lines, respectively. 
Table \ref{tab:bins} presents the $r_{gc}$ intervals defined to draw the anisotropy profiles, the velocity dispersion, corresponding $\beta_s$, and $\beta$ for both complete and clean samples. These median values are used to present the profiles in Figures \ref{fig:vel_disp} and \ref{fig:beta}. The parameters obtained from segmented linear regression applied to the profiles are summarized in Table \ref{tab:slop}.

\begin{table}[h]
\caption{Intervals of Galactocentric distance ($r_{gc}$), velocity dispersions ($\sigma_i$, in $\rm km \, s^{-1}$), velocity anisotropy parameter ($\beta$), and velocity anisotropy index ($\beta_s$) for complete and clean samples.} \label{tab:bins}
\centering
\begin{tabular}{c|ccccc|ccccc}
\hline \hline 
\multirow{2}{1.6cm}{$r_{gc} \ \rm (kpc)$} &
  \multicolumn{5}{c|}{Complete sample} &
  \multicolumn{5}{c}{Clean sample} \\
   & $\sigma_r$ & $\sigma_\theta$ & $\sigma_\phi$ & $\beta_s$ & $\beta$ & $\sigma_r$ & $\sigma_\theta$ & $\sigma_\phi$ & $\beta_s$ & $\beta$ \\
\hline
(2, 6{]}   & 146.11 & 101.87 & 120.40 & 0.27  & 0.42  & 121.62 & 86.58  & 99.16  & 0.26  & 0.41  \\
(6, 7{]}   & 131.99 & 90.22  & 105.82 & 0.29  & 0.45  & 94.74  & 81.63  & 84.84  & 0.13  & 0.24  \\
(7, 8{]}   & 142.78 & 94.08  & 106.02 & 0.34  & 0.51  & 111.96 & 75.36  & 83.10  & 0.34  & 0.51  \\
(8, 9{]}   & 141.81 & 92.63  & 97.56  & 0.38  & 0.55  & 116.64 & 82.09  & 85.79  & 0.32  & 0.48  \\
(9, 10{]}  & 147.06 & 88.38  & 94.94  & 0.44  & 0.61  & 146.04 & 96.96  & 102.80 & 0.36  & 0.53  \\
(10, 12{]} & 146.33 & 88.85  & 86.50  & 0.47  & 0.64  & 151.34 & 100.11 & 103.60 & 0.37  & 0.55  \\
(12, 14{]} & 138.16 & 86.52  & 85.40  & 0.44  & 0.61  & 151.88 & 103.46 & 108.59 & 0.34  & 0.51  \\
(14, 16{]} & 130.17 & 85.21  & 86.59  & 0.40  & 0.57  & 147.87 & 94.93  & 102.50 & 0.38  & 0.55  \\
(16, 18{]} & 121.00 & 85.01  & 81.25  & 0.36  & 0.53  & 146.09 & 92.10  & 92.26  & 0.44  & 0.61  \\
(18, 20{]} & 115.90 & 80.38  & 79.61  & 0.36  & 0.53  & 145.28 & 90.23  & 89.37  & 0.44  & 0.61  \\
(20, 22{]} & 111.87 & 177.53 & 182.78 & 0.32  & 0.49  & 138.14 & 186.53 & 193.73 & 0.41  & 0.58  \\
(22, 24{]} & 105.91 & 180.31 & 181.99 & 0.26  & 0.41  & 130.26 & 182.56 & 186.02 & 0.40  & 0.58  \\
(24, 26{]} & 104.02 & 184.72 & 181.08 & 0.23  & 0.37  & 124.12 & 177.08 & 177.36 & 0.43  & 0.60  \\
(26, 28{]} & 105.95 & 188.41 & 183.09 & 0.21  & 0.35  & 121.92 & 175.09 & 169.06 & 0.49  & 0.65  \\
(28, 31{]} & 108.68 & 198.82 & 183.90 & 0.17  & 0.29  & 114.50 & 176.41 & 161.78 & 0.46  & 0.63  \\
(31, 34{]} & 111.69 & 105.43 & 188.36 & 0.13  & 0.23  & 105.45 & 175.66 & 161.62 & 0.39  & 0.56  \\
(34, 37{]} & 103.10 & 110.95 & 100.00 & -0.02 & -0.04 & 193.05 & 174.81 & 165.60 & 0.27  & 0.42  \\
(37, 40{]} & 199.72 & 117.12 & 109.61 & -0.13 & -0.30 & 194.54 & 180.87 & 170.03 & 0.22  & 0.36  \\
(40, 45{]} & 101.36 & 132.41 & 123.05 & -0.22 & -0.58 & 198.15 & 198.70 & 176.21 & 0.11  & 0.20  \\
(45, 80{]} & 191.74 & 148.95 & 147.03 & -0.44 & -1.60 & 188.97 & 125.82 & 185.23 & -0.18 & -0.43 \\
\hline \hline 
\end{tabular}
\end{table}

\begin{table}[h]
\caption{Slopes ($a_i$, in $\rm km \, s^{-1} \, kpc$), breakpoints ($k_i$, in kpc), and respective standard deviations obtained with a segmented linear regression for the complete and clean samples in the velocity dispersion ($\sigma_r$, $\sigma_\theta$ and $\sigma_\phi$) and anisotropy ($\beta_s$) profiles.} \label{tab:slop}
\centering
\begin{tabular}{cccc}
 $\sigma_r$ & $\sigma_\theta$ & $\sigma_\phi$ & $\beta_s$ \\
\hline \hline 
\multicolumn{4}{c}{Complete} \\ 
\hline 
\multirow{7}{3cm}{$a_1 = -2.22 \pm 0.30$ \\ $k_1 = 22.7 \pm 2.9$ \\ $a_2 = -0.53 \pm 0.24$} & \multirow{7}{3cm}{$a_1 = -1.16 \pm 0.17$ \\ $k_1 = 21.4 \pm 0.7$ \\ $a_2 = 2.36 \pm 0.10$} & \multirow{7}{3cm}{$a_1 = -5.52 \pm 0.57$ \\ $k_1 = 11.2 \pm 0.5$ \\ $a_2 = -0.23 \pm 0.20$ \\ $k_2 = 28.5 \pm 0.9$ \\ $a_3 = -2.78 \pm 0.14$} & $a_1 = 0.042 \pm 0.004$\\
 & & & $k_1 = 10.7 \pm 0.3$ \\
 & & & $a_2 = -0.017 \pm 0.001$ \\
 & & & $k_2 = 32.9 \pm 0.6$ \\
 & & & $a_3 = -0.044 \pm 0.003$ \\
 & & & $k_3 = 38.8 \pm 1.0$ \\
 & & & $a_4 = -0.020 \pm 0.002$ \\
\hline 
\multicolumn{4}{c}{Clean} \\ 
\hline
$a_1 = -20.8 \pm 5.2$ & $a_1 = -4.9 \pm 1.9$ & $a_1 = -11.1 \pm 4.3$ & \multirow{7}{3cm}{$a_1 = 0.010 \pm 0.002$ \\ $k_1 = 28.1 \pm 1.2$ \\ $a_2 = -0.027 \pm 0.003$ } \\  
$k_1 = 6.6 \pm 0.2$ & $k_1 = 7.9 \pm 0.3$ & $k_1 = 7.0 \pm 0.4$ &  \\
$a_2 = 17.2 \pm 3.4$ & $a_2 = 14.8 \pm 4.3$ & $a_2 = 6.7 \pm 1.5$ & \\   
$k_2 = 10.5 \pm 0.4$ & $k_2 = 9.9 \pm 0.3$ & $k_2 = 11.7 \pm 0.6$ & \\
$a_3 = -2.4 \pm 0.2$ & $a_3 = -1.4 \pm 0.1$ & $a_3 = -2.6 \pm 0.2$ & \\ 
$k_3 = 35.8 \pm 2.9$ & $k_3 = 33.6 \pm 0.7$ & $k_3 = 30.9 \pm 1.1$ & \\ 
$a_4 = -0.49 \pm 0.46$ & $a_4 = 3.02 \pm 0.23$ & $a_4 = 1.17 \pm 0.25$ & \\ 
\hline \hline 
\end{tabular}
\end{table}


\bibliography{refs}{}
\bibliographystyle{aasjournalv7}



\end{document}